\documentclass[aps,twocolumn,pra,nofootinbib]{revtex4}
\usepackage{amsmath,amssymb,bm}
\usepackage{graphicx}
\usepackage{epstopdf}
\usepackage{latexsym}
\usepackage{subfigure}
\usepackage[usenames,dvipsnames]{color}
\usepackage{hyperref}
\usepackage{natbib}
\usepackage{moreverb}
\begin{document}

\newcommand{\ns}{\mathcal{N}_{\mathrm{s}}}

\title{Simulating fermions in spin-dependent potentials with spin models on an energy lattice}
\date{\today}
\author{Michael L. Wall}
\affiliation{JILA, NIST and University of Colorado, 440 UCB, Boulder, CO 80309, USA}
\affiliation{The Johns Hopkins University Applied Physics Laboratory, Laurel, MD, 20723, USA}

\newcommand{\red}[1]{{ \bf \color{red}#1}}

\begin{abstract}
We study spin-1/2 fermions in spin dependent potentials under the \emph{spin model approximation}, in which interatomic collisions that change the total occupation of single-particle modes are ignored.   The spin model approximation maps the interacting fermion problem to an ensemble of lattice spin models in energy space, where spin-spin interactions are long-ranged and spin-anisotropic.  We show that the spin model approximation is accurate for weak interactions compared to the harmonic oscillator frequency, and captures the collective spin dynamics to timescales much longer than would be expected from perturbation theory.  We explore corrections to the spin model, and the relative importance of corrections when realistic anharmonic potential corrections are taken into account.  Additionally, we present numerical techniques that are useful for analysis of spin models on an energy lattice, including enacting a change of single-particle basis on a many-body state as an effective time evolution, and fitting of spatially inhomogeneous long-range interactions with exponentials.  This latter technique is useful for constructing matrix product operators for use in DMRG analyses, and may have broader applicability within the tensor network community.
\end{abstract}

\maketitle

\section{Introduction}

Lattice models of quantum magnetism, which describe the interactions between quantum mechanical spins on a regular array, are the oldest and still most prevalent realizations of strongly correlated quantum many-body systems~\cite{Auerbach_94}.  While realizations of quantum magnetism occur in a variety of solid state and other condensed matter settings, modern advances in atomic, molecular, and optical (AMO) systems have opened the possibility for engineering models in which the dimensionality, strength, and even range of spin-spin interactions are all amenable to experimental control~\cite{simon2011quantum,britton2012engineered,islam2013emergence,hart2015observation,PhysRevLett.115.260401}.  In addition, AMO experiments have at hand unique tools for microscopically probing and characterizing magnetic order~\cite{bakr2009quantum,PhysRevA.92.063633}.  In many AMO platforms, interactions are short range and spin-independent, and so quantum magnetism arises from a combination of atomic motion in a trap and quantum statistics through, e.g., the superexchange mechanism~\cite{PhysRev.79.350,Trotzky18012008,Brown01052015} or through direct spin exchange~\cite{PhysRevLett.98.070501,anderlini2007controlled,kaufman2015entangling}.  Because of this complex interplay, properly understanding magnetic phenomena in AMO systems requires an approach which treats motion, interactions, and quantum statistics on a common footing.  

 \begin{figure}[t]
\includegraphics[width=.8\columnwidth,angle=0]{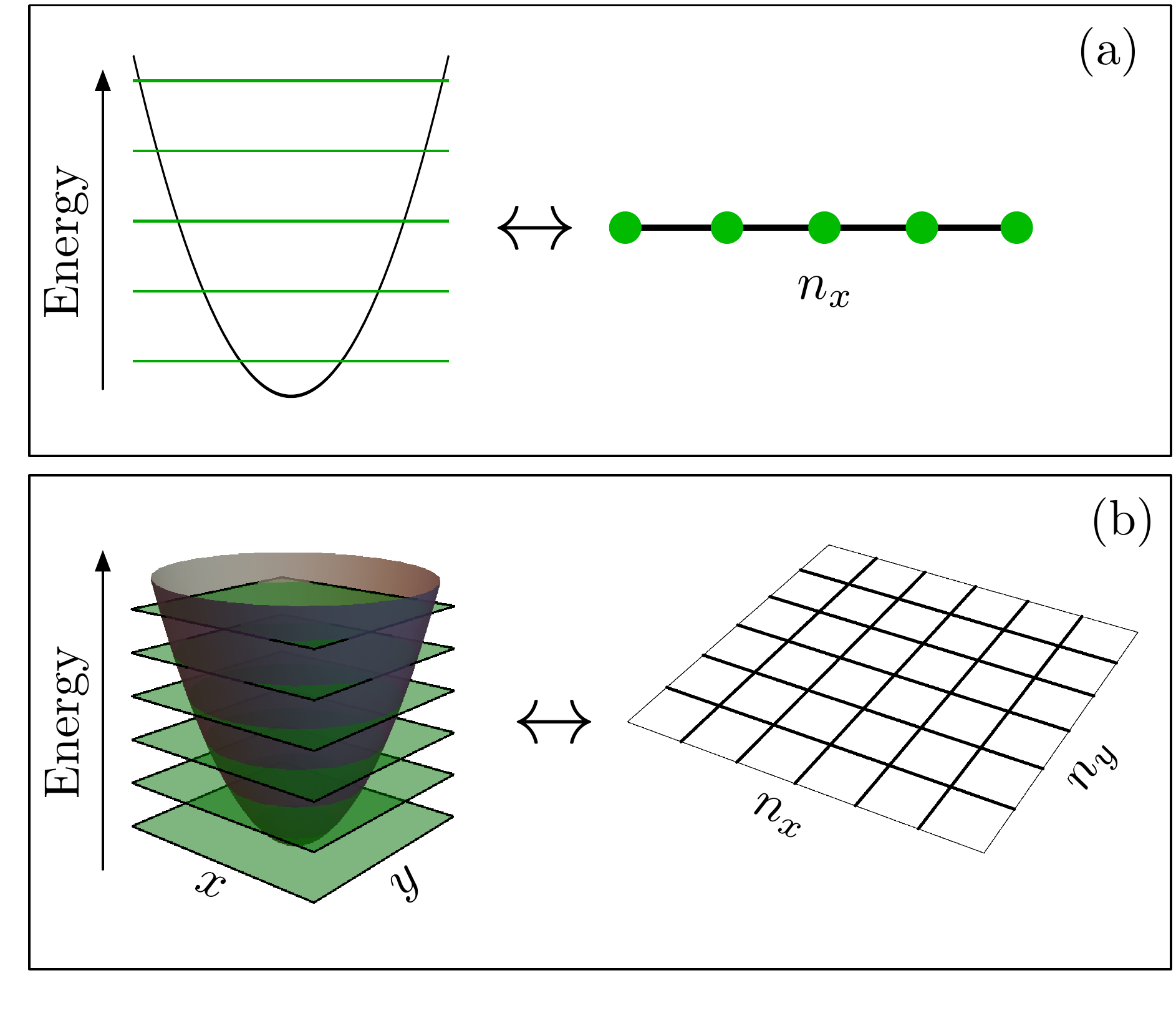}
\caption{(color online) \emph{Energy Lattices.}  The single-particle quantum numbers of a trap can be thought of as enumerating the sites of a regular lattice in energy space, which we call an energy lattice.  For $D$-dimensional harmonic traps, the number of quanta along each Cartesian direction enumerates a $D$-dimensional cubic energy lattice.  Examples are given for (a) a 1D Harmonic trap and (b) a 2D Harmonic trap.
 \label{fig:spinmodel}}
\end{figure}

In this work, we explore such an approach which is applicable to weakly interacting two-component particles in tight traps.  The key approximation in our approach, whose validity we investigate in detail, is that the single-particle motional states are not changed by interactions, but interactions can affect spin dynamics though exchange.  While this approach is only formally valid in the regime in which the characteristic interaction energy per particle $U$ is weak compared to the trap energy scale $\hbar \omega$, many relevant quantities, such as the demagnetization time of a polarized ensemble, can be evaluated with much greater accuracy~\cite{koller14}.  If the single-particle mode quantum numbers of the trap are thought of as enumerating positions on a regular spatial array which we call the \emph{energy lattice} (Fig.~\ref{fig:spinmodel}), then the Hamiltonian with this approximation takes the form of a spin model, and hence this is known as the \emph{spin model approximation}~\cite{Martin09082013,rey14}.  In contrast to most spin models formulated in real space, spin models on the energy lattice feature long-range and inhomogeneous spin-spin interactions and effective magnetic fields.  Additionally, while the interaction Hamiltonian with this approximation takes the form of a spin model, the Hilbert space it acts upon is still that of spin-1/2 fermions, and so the single-particle dynamics of arbitrarily correlated motional states can be exactly accounted for.

The key advantage of the above prescription is that it provides a description of the quantum dynamics of interacting fermions model in terms of a spin model, as tools for simulating the dynamics of quantum spin models are, in general, better developed than their fermionic counterparts~\cite{PhysRevX.5.011022}.  The resulting spin models have spin-spin interactions which are long-ranged, which enables analytical understanding through the use of collective or nearly-collective models~\cite{PhysRevA.77.052305,pegahan2019spin}.  In addition, as shown in Ref.~\cite{PhysRevLett.117.195302}, the combined effects of motion and spin dynamics in the lowest order can be captured within the spin model approximation by simulating a collection of spin models, each defined on a different energy lattice.  These different energy lattices, each of which has its own set of spin model couplings, represent different motional configurations which are involved in the dynamics.  Hence, the coherences between different energy lattice dynamics carry information about the interplay of spin and motion.

In addition to exploring the foundations and validity of the spin model approach, we also present numerical techniques that are useful when dealing with dynamics on the energy lattice.  In particular, we present an account of methods that were used in Ref.~\cite{PhysRevLett.117.195302} to study spin model dynamics with matrix product state (MPS) simulations.  MPSs, which are the underlying variational framework of the density-matrix renormalization group (DMRG) algorithm~\cite{White1992,White93}, have a long history in quantum spin systems~\cite{PhysRevB.48.3844,fannes1992}, and continue to be important in the frontiers of quantum magnetism, including in long-range spin systems~\cite{PhysRevX.3.031015} and in higher dimensions~\cite{Depenbrock}.  While some of our methodologies are specific to MPSs, others, such as global transformations between energy lattice representations, have broader applicability.  Some of the MPS tools we develop to study the present spin models, such as the representation of long-range, non-translationally invariant interactions, may also have broader applicability in other DMRG applications, such as in MPS-based approaches to quantum chemistry~\cite{doi:10.1146/annurev-physchem-032210-103338,qchemMPO}.  In addition, the spin model approximation, which we only rigorously benchmark in 1D (see also Ref.~\cite{koller14}), is expected to be valid in higher dimensions, a claim supported by recent 3D experiments~\cite{smale2019observation}.  This will enable the use of approximate techniques for spin systems, such as the truncated Wigner approximation~\cite{PhysRevX.5.011022,Polkovnikov20101790} to study the out-of-equilibrium dynamics of fermions in dimensions greater than one, where no efficient, unbiased algorithms currently exist.

\section{States, interactions, and observables on the energy lattice}
\label{eq:EnergyLatt}

A key component of our approach is that of an \emph{energy lattice}, which is a discrete set of points indexed by single-particle energy.  While there can be considerable freedom in choosing how the lattice sites are arranged and indexed, for (near-)harmonic traps it is natural to choose the energy lattice to be a regular cubic lattice with the same dimensionality as the trapped system, as in Fig.~\ref{fig:spinmodel}.  Throughout, we will denote the trap mode index with Roman letters $n$, with the understanding that this can be immediately generalized to a vector of indices $\mathbf{n}$ in higher-dimensional scenarios, and the spin state with greek letters $\sigma$.  

Given a set of these possibly spin-dependent trap states $\{\psi_{n\sigma}\left(x\right)\}$ which spans the low-energy part of Hilbert space that is of interest to us, we can expand the field operator $\hat{\psi}\left(x\right)$ in this set.  Then, substituting this expansion into the second-quantized representation of the many-body Hamiltonian generates an effective low-energy model, in analogy with the procedure commonly used to derive lattice Hubbard models using Wannier functions~\cite{Jaksch1998,lewensteinM2007}.  For spin-1/2 fermions experiencing $s$- and $p$-wave interactions modeled by contact pseduopotentials~\cite{Idziaszek_Calarco_06}, the resulting many-body model is
\begin{align}
\label{eq:FullH} \hat{H}&=\sum_{n,\sigma} \hbar \omega_{n\sigma} \hat{n}_{n,\sigma}\\
\nonumber &+\frac{1}{2}\sum_{n_1,n_2; n_2',n_1'}\sum_{\sigma \sigma'}I_{n_1,n_2; n_2',n_1'}^{\sigma\sigma'} \hat{a}^{\dagger}_{n_1\sigma}\hat{a}^{\dagger}_{n_2\sigma'}\hat{a}_{n_2'\sigma'}\hat{a}_{n_1'\sigma}
\end{align}
where the first line is the single-particle energy of trap state $|n\rangle$ in spin state $\sigma$ and the second line represents the interaction Hamiltonian.  In this expression, $\hat{a}_{n\sigma}$ destroys a particle in mode $n$ and spin state $\sigma$, $\hat{n}_{n,\sigma}=\hat{a}^{\dagger}_{n\sigma}\hat{a}_{n\sigma}$, and the interaction matrix elements are
\begin{align}
\nonumber I_{n_1,n_2; n_2',n_1'}^{\sigma\sigma'}&=\frac{4\pi \hbar^2 a_s\left(1-\delta_{\sigma\sigma'}\right)}{M}S_{n_1,n_2; n_2',n_1'}^{\sigma\sigma'}\\
&+\frac{6\pi \hbar^2b_{\sigma\sigma'}^3}{M}P_{n_1,n_2; n_2',n_1'}^{\sigma\sigma'}
\end{align}
where $a_s$ is the $s$-wave scattering length, $b_{\sigma\sigma'}^3$ are the $p$-wave scattering volumes, and the geometrical integrals $S$ and $P$ are given as
\begin{align}
\label{eq:swavepp} &S_{n_1,n_2; n_2',n_1'}^{\sigma\sigma'}=\int d\mathbf{r}\psi_{n_1\sigma}^{\star}\left(\mathbf{r}\right)\psi_{n_2\sigma'}^{\star}\left(\mathbf{r}\right)\psi_{n_2'\sigma'}\left(\mathbf{r}\right)\psi_{n_1'\sigma}\left(\mathbf{r}\right)\, ,\\
\nonumber  &P_{n_1,n_2; n_2',n_1'}^{\sigma\sigma'}=\int d\mathbf{r}\\
\nonumber & \times  \left[\psi_{n_1\sigma}^{\star}\left(\mathbf{r}\right)\left(\nabla \psi_{n_2\sigma'}^{\star}\left(\mathbf{r}\right)\right)-\left(\nabla \psi_{n_1\sigma}^{\star}\left(\mathbf{r}\right)\right)\psi_{n_2\sigma'}^{\star}\left(\mathbf{r}\right)\right]\\
\label{eq:pwavepp}&\cdot \left[\psi_{n_1'\sigma}\left(\mathbf{r}\right)\left(\nabla \psi_{n_2'\sigma'}\left(\mathbf{r}\right)\right)-\left(\nabla \psi_{n_1'\sigma}\left(\mathbf{r}\right)\right)\psi_{n_2'\sigma'}\left(\mathbf{r}\right)\right]\, .
\end{align}
We use discrete variable representations for the representation and manipulation of position-space wavefunctions in this work, including the evaluation of these integrals, because of their simplicity, flexibility, and exponential convergence~\cite{PhysRevA.92.013610}. 

Translating observables to the energy lattice representation also proceeds straightforwardly by using the expansion of the field operator in terms of the trap states.  For example, the density of the collective spin raising operator $\hat{\mathcal{S}}^+\left(x\right)$ can be written as
\begin{align}
\hat{\mathcal{S}}^+\left(x\right)&=\sum_{n m} \psi_{n\uparrow}\left(x\right)\psi_{m\downarrow}\left(x\right) \hat{a}^{\dagger}_{n\uparrow}\hat{a}_{m\downarrow}\, .
\end{align}
Integrating this density, we find the collective raising operator as
\begin{align}
\hat{\mathcal{S}}^+&=\sum_{n m} \gamma_{nm} \hat{a}^{\dagger}_{n\uparrow}\hat{a}_{m\downarrow}\, ,
\end{align}
where $\gamma_{nm}=\int dx \psi_{n\uparrow}\left(x\right)\psi_{m \downarrow}\left(x\right)$.  For spin-independent traps, $\gamma_{nm}=\delta_{nm}$, but for spin-dependent traps this matrix is non-diagonal, with its non-zero elements having subunity modulus.  Hence, coherences between different positions on the energy lattice are important for capturing the spatial dependence of the fermionic spin density.

The spin model approximation~\cite{Martin09082013,rey14} consists of restricting the single-particle modes to a fixed set, and only allowing for interaction terms $I_{n'm';m n}^{\sigma\sigma'}$ which do not change the single-particle modes participating in the collision.  Namely, this approximation keeps the direct $I_{nm;mn}^{\sigma\sigma'}$ and exchange $I_{mn;mn}^{\sigma\sigma'}$ terms.  With this restriction, the Hamiltonian Eq.~\eqref{eq:FullH} reduces to
\begin{align}
\nonumber \hat{H}&=\sum_{n,\sigma} \hbar \omega_{n\sigma} \hat{n}_{n,\sigma}+\sum_{n}I_{nnnn}^{\uparrow\downarrow}\hat{n}_{n\uparrow}\hat{n}_{n\downarrow}+\sum_{n\ne n'}K_{nn'}\hat{n}_{n}\hat{n}_{n'}\\
\nonumber &+\sum_{n\ne n'}[J^{\perp}_{nn'}\left(\hat{S}^X_n\hat{S}^X_{n'}+\hat{S}^Y_n\hat{S}^Y_{n'}\right)+J^Z_{nn'}\hat{S}^Z_n\hat{S}^Z_{n'}\\
\label{eq:genSM}&+C_{nn'}\hat{S}^Z_n\hat{n}_{n'}+D_{nn'}\hat{n}_n\hat{S}^Z_{n'}]
\end{align}
The general form of the spin model is that of a long-range XXZ model in with inhomogeneous longitudinal fields (the terms with $C$ and $D$ coefficients).  Here, the parameters appearing in the spin model are defined as
\begin{align}
\nonumber K_{nn'}&=\frac{1}{2}(2I^{\uparrow\uparrow}_{nn'n'n}+2I^{\downarrow\downarrow}_{nn'n'n}+I^{\uparrow\downarrow}_{nn'n'n}+I^{\downarrow\uparrow}_{nn'n'n})\, ,\\
J^{\perp}_{nn'}&=-I^{\uparrow\downarrow}_{nn'nn'}\, ,\\
\nonumber J^Z_{nn'}&=\frac{1}{2}(2I^{\uparrow\uparrow}_{nn'n'n}+2I^{\downarrow\downarrow}_{nn'n'n}-I^{\uparrow\downarrow}_{nn'n'n}-I^{\downarrow\uparrow}_{nn'n'n})\, ,\\
\nonumber C_{nn'}&=\frac{1}{2}(2I^{\uparrow\uparrow}_{nn'n'n}-2I^{\downarrow\downarrow}_{nn'n'n}-I^{\uparrow\downarrow}_{nn'n'n}+I^{\downarrow\uparrow}_{nn'n'n})\, ,\\
\nonumber D_{nn'}&=\frac{1}{2}(2I^{\uparrow\uparrow}_{nn'n'n}-2I^{\downarrow\downarrow}_{nn'n'n}+I^{\uparrow\downarrow}_{nn'n'n}-I^{\downarrow\uparrow}_{nn'n'n})\, ,
\end{align}
where we have used the fact that $I^{\uparrow\downarrow}_{nn'nn'}=I^{\downarrow\uparrow}_{nn'nn'}$ and $I^{\sigma\sigma}_{nn'nn'}=-I^{\sigma\sigma}_{nn'n'n}$.  We note that the coefficients $C$ and $D$ are only nonzero in the case that the trap is spin-dependent.  In the context of optical lattice clocks where high precision measurements are possible and temperatures correspond to a thermally averaged number of trap quanta $\sim50$~\cite{Martin09082013}, both $s$- and $p$-wave collisions play an important role in the spin model.  However, in ultracold or lower-precision scenarios, we can neglect the $p$-wave components and keep only the $s$-wave, resulting in the simpler coefficients
\begin{align}
 \mathcal{K}_{nn'}&=\frac{1}{2}(I^{\uparrow\downarrow}_{nn'n'n}+I^{\downarrow\uparrow}_{nn'n'n})\, ,\\
\label{eq:sJperp}\mathcal{J}^{\perp}_{nn'}&=-I^{\uparrow\downarrow}_{nn'nn'}\, ,\\
\mathcal{J}^Z_{nn'}&=-\frac{1}{2}(I^{\uparrow\downarrow}_{nn'n'n}+I^{\downarrow\uparrow}_{nn'n'n})\, ,\\
\label{eq:sC}\mathcal{C}_{nn'}&=\frac{1}{2}(I^{\downarrow\uparrow}_{nn'n'n}-I^{\uparrow\downarrow}_{nn'n'n})\, ,\\
 \mathcal{D}_{nn'}&=\frac{1}{2}(I^{\uparrow\downarrow}_{nn'n'n}-I^{\downarrow\uparrow}_{nn'n'n})\, .
\end{align}

Before proceeding, we would also like to note that the essential idea of the spin model, which is to restrict the Hamiltonian to energy-conserving direct and exchange processes, is quite old and has appeared in many contexts.  Perhaps the earliest relevant use of this approach for studying quantum effects in collisions of spinful particles is from Lhuillier and Lalo\"{e}~\cite{refId0,refId2}.  Many applications of the spin model idea to cold atomic systems arrived in the 2000s, following experiments at JILA~\cite{lewandowski02} and Duke~\cite{duke08}, which focused on fully collective spin systems~\cite{PhysRevLett.88.230403,PhysRevLett.104.010801,PhysRevLett.103.010401}, kinetic theory approaches~\cite{PhysRevLett.88.230404,PhysRevLett.102.215301}, and, later, on interaction effects in atomic clocks~\cite{PhysRevLett.103.260402,PhysRevLett.103.113202}.  The spin model and related approaches have been applied to the recent experiments reported in Ref.~\cite{pegahan2019spin,smale2019observation}.

Perhaps the most significant difference of our approach compared to those listed above are that we only enact the spin model approximation at the level of the Hamiltonian, Eq.~\eqref{eq:genSM}.  In particular, our spin model framework places no restrictions whatsoever on the state, which is still defined on the Hilbert space of spin-1/2 fermions in general, and so can have arbitrary correlations between spins or between spin and motional degrees of freedom.  This fact is essential for applications to spin-dependent quantum quenches~\cite{PhysRevLett.117.195302}.  The other significant difference with many of the works above is that our approach is fully quantum; that is, we do not study the spin model within a mean-field or semiclassical approach, but instead use tools developed for strongly correlated quantum systems.

\subsection{Behavior of the spin model parameters for spin-dependent traps}

 \begin{figure*}[p]
\includegraphics[width=1.8\columnwidth]{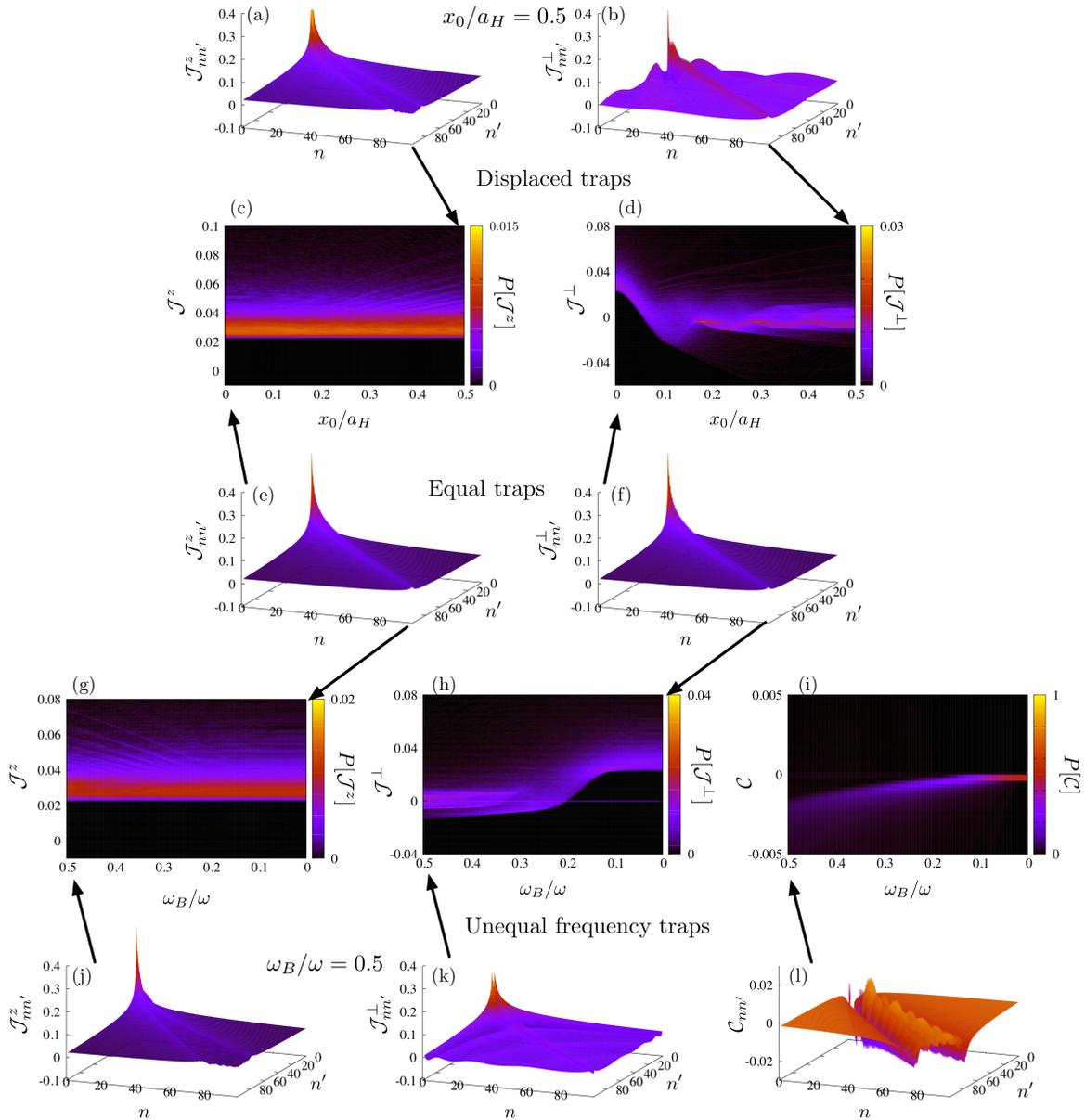}
\caption{(color online) \emph{Effective spin-spin interactions on the energy lattice.}  Interaction parameters for (a)-(d) harmonic traps with a spin-dependent displacement $\pm x_0$, (e)-(f) spin-independent harmonic traps, and (g)-(l) harmonic traps with a spin-dependent harmonic trapping frequency $\sqrt{\omega^2\pm \omega_B^2}$.  Panels (a), (b), (e), (f), (j), (k), and (l) are surface plots of interaction parameters as functions of the sites of the energy lattice, showing the long-range, non-translationally invariant character of the interactions.  Additionally, we see that introducing spin dependence to the trap affects the direct interactions $\mathcal{J}^z$ generally only slightly, but has a drastic effect upon the exchange coefficients $\mathcal{J}^z$, reducing their magnitude and introducing negative values.  This is also seen in panels (c), (d), (g), (h), and (i), which are histograms giving the probability $P\left[\mathcal{J}\right]$ of obtaining the value $\mathcal{J}$ for an interaction parameter in the first 50 modes as functions of the trap spin dependence.
 \label{fig:Interactions}}
\end{figure*}

Plots of the $s$-wave spin model parameters Eq.~\eqref{eq:sJperp}-\eqref{eq:sC} for a 1D harmonic trap are given in Fig.~\ref{fig:Interactions}.  Here and throughout this work, these parameters are measured in units of the $s$-wave interaction strength $U\equiv 4\pi \hbar^2 a_s^{(\mathrm{1D})}/(M a_H)$, with $a_s^{(\mathrm{1D})}$ the effective 1D scattering length obtained by integrating along tight confinement in the transverse directions\footnote{Note that $a_s^{(\mathrm{1D})}$ has units of inverse length.} .  We will first look at panels (e) and (f), which are plots of $\mathcal{J}^z_{nn'}$ and $\mathcal{J}^{\perp}_{nn'}$ for spin-independent traps ($\mathcal{C}_{nn'}=\mathcal{D}_{nn'}=0$ for spin-independent traps).  First, we note that interactions between different modes on the energy lattice are long ranged even though the underlying interactions are short-ranged in real space due to the fact that the single-particle motional states have significant spatial overlap.  A rough estimate of the asymptotic decay of these interactions for spin-independent traps is $1/\sqrt{n-m}$.  Next, we note that the interactions are not translationally invariant on the energy lattice, which is to say that, e.g., $\mathcal{J}^z_{nn'}\ne f(|n-n'|)$, in contrast to interactions in real space.

For spin-independent traps, $\mathcal{J}^z_{nn'}=\mathcal{J}^{\perp}_{nn'}$, resulting in a Heisenberg model with SU(2) spin-rotation symmetry; this reflects the underlying symmetry of $s$-wave interactions.  When the trap becomes spin-dependent this is no longer the case, as we have explicitly broken the symmetry between spin states.  Fig.~\ref{fig:Interactions} shows the interaction parameters computed for two special cases of spin-dependent traps.  The first, corresponding to the upper panels (a)-(d), are for a trap with a spin-dependent displacement and the second, corresponding to the lower panels (g)-(l), are for a trap with a spin-dependent trapping frequency.  The eigenstates for the former are shifted harmonic oscillator states, $\psi_{\sigma n}\left(x\right)=\phi_{n}\left(x+\sigma \frac{x_0}{a_H}\right)$, with $\phi_n\left(x\right)$ the harmonic oscillator eigenstates, $\sigma=+/-$ for $\uparrow/\downarrow$, $x_0$ the displacement of the traps, and $a_H$ the harmonic oscillator length, and the states for the latter are $\psi_{\sigma n}\left(x\right)=(1+\sigma\frac{\omega_B^2}{\omega^2})^{1/8} \phi_n(x(1+\sigma\frac{\omega_B^2}{\omega^2})^{1/4})$, corresponding to spin-dependent trapping frequencies $\omega_{\sigma}=\sqrt{\omega^2+\sigma\omega_B^2}$.

From Eq.~\eqref{eq:swavepp}, we see that the parameters $\mathcal{J}^Z$ involve the integral of a product of two densities, and hence are positive definite.  On the other hand, the exchange coefficients $\mathcal{J}^{\perp}$ involve integrating a product of terms of the form $\psi_{\sigma n}(x)\psi_{\sigma n'}(x)$, which themselves integrate to zero and so are not positive definite.  As the traps for spin up and spin down are displaced in opposite directions, the $\mathcal{J}^Z$s generally decrease in magnitude, but this decrease is small for shifts small compared to a harmonic oscillator length.  This is shown in Fig.~\ref{fig:Interactions}(c), which is a histogram of the $\mathcal{J}^{\perp}$s for the first 50 modes as a function of the spin-dependent displacements.  On the other hand, as shown in Fig.~\ref{fig:Interactions}(d), the exchange coefficients decay much more quickly with displacement, and also take on negative values.  Snapshots of the complete mode dependence for the largest displacement are given in panels (a) and (b); the oscillatory behavior of $\mathcal{J}^{\perp}$ with mode number is clearly visible.  We note that, because the traps have the same shape but are simply displaced, $\mathcal{C}$ and $\mathcal{D}$ remain zero for any $x_0$.

We now turn to the traps with spin-dependent frequency $\omega_{\sigma}=\sqrt{\omega^2+\sigma\omega_B^2}$, shown in Fig.~\ref{fig:Interactions}(g)-(l).  Similar to the displaced traps in panels (c) and (d), we see that the direct interactions $\mathcal{J}^Z$ are relatively insensitive (statistically) to the spin-dependence of the trap, while spin dependence causes the exchange terms to drop in magnitude and take on negative values.  In addition, the $\mathcal{C}=-\mathcal{D}$ terms (panel (i)) become nonzero in this case, but are generally quite small.  Finally, panels (j)-(l) show a snapshot of the interaction parameters for the largest disparity in frequencies.

\section{Numerical techniques}

In this section we present two methods that are useful for simulating spin systems on an energy lattice.  The first is to recast global transformations of a state between two energy lattices, as occurs when the single-particle potential is abruptly modified, in terms of long-range time evolution under a non-interacting Hamiltonian.  Here, we benchmark our methods against two cases where the transformation is known analytically, namely, harmonic oscillators subject to sudden spin-dependent displacement or change in frequency.  The second numerical tool we provide is to represent non-translationally invariant interactions, such as occur on the energy lattice, in terms of sums of exponentials with site-dependent weighting and exponential decay coefficients.  Such a representation is essential for building matrix product operator (MPO) representations of spin model Hamiltonians to use in MPS calculations.  We will not cover the basics of MPS calculations here, as there are many excellent reviews devoted to the subject~\cite{Schollwoeck,Orus}.  More information about MPOs and how they can be used to enhance MPS algorithms may be found in Refs.~\cite{PhysRevLett.93.207204,mcculloch2007density,PhysRevA.81.062337,Wall_Carr_12}.  We note that our numerical methodologies are by no means restricted to MPS algorithms, but may be used anywhere global transformations or inhomogeneous interactions may be found.  In addition, within the MPS/DMRG community, these methods may find uses in other model applications, e.g., quantum chemistry~\cite{doi:10.1146/annurev-physchem-032210-103338,qchemMPO}.

\subsection{Transforming between global representations}
\label{sec:GlobalTransform}
A ubiquitous approach in ultracold gas experiments is to probe a system by suddenly quenching some parameter in the Hamiltonian and observing the ensuing non-equilibrium dynamics; this is the basis of Ramsey spectroscopy, for example.  When the quenched parameter is related to the trapping potential, the dynamics within the spin model approach is calculated by projecting the initial state onto a new energy lattice defined by the quenched trap and then evolving the spin dynamics with Eq.~\eqref{eq:genSM} while keeping the mode occupations in the new trap fixed.  The transformation between states on different energy lattices is generally a very complex, highly non-local operation on the energy lattice.  However, as we show here, there is a well-defined procedure by which we can cast this quench as a time evolution under a simulated ``Hamiltonian" which consists of long-range and inhomogeneous free-particle hopping.  Hence, this quench procedure can be simulated numerically using any method capable of long-range time evolution.  An explicit matrix product operator (MPO) form for this inhomogeneous, long-range Hamiltonian, which is useful for time evolution within the matrix product state (MPS) formalism, can be obtained using the methods of the next section.  Many MPS-based methods for time evolution under long-range Hamiltonians exist, including Krylov subspace methods~\cite{Manmana,GarciaRipoll,Wall_Carr_12}, the time-dependent variational principle~\cite{haegeman2014unifying}, and the local Runge-Kutta method~\cite{Zaletel} used in Ref.~\cite{PhysRevLett.117.195302}.  A nice feature of our effective time-evolution approach is that this change of basis can be applied to any arbitrary state; it is not restricted to product states.

A transformation between the two single-particle bases $\{|n\rangle\}$ and $\{|\tilde{n}\rangle\}$ is provided by the matrix
\begin{align}
\label{eq:Unntilde} U_{{n}\tilde{{n}}}&=\langle {n}|\tilde{{n}}\rangle \, .
\end{align}
As we let the number of states $|n\rangle$ and $|\tilde{n}\rangle$ tend to infinity, this becomes a unitary matrix.  Postponing temporarily questions about the unitarity of this matrix for finite-dimensional representations, we note that this unitary operator can be written as the exponential of a Hermitian operator $\hat{U}=\exp(-i \hat{H})$, which has the interpretation of time evolution for unit time with an ersatz ``Hamiltonian," which we will call the change of basis Hamiltonian (COBH).  Further, as the COBH generates a transformation between single-particle Hilbert spaces, its second-quantized counterpart is a non-interacting Hamiltonian with, in general, long-range and inhomogeneous hopping.

There are two important situations in which the COBH can be determined exactly, which are the cases of a harmonic oscillator subject to a sudden shift in trap center and a sudden change of trap frequency.  In the case of a shifted oscillator, the unitary transformation may be written as
\begin{align}
\hat{U}_{\mathrm{shifted}}&=e^{\lambda \left(\hat{a}-\hat{a}^{\dagger}\right)/\sqrt{2}}\, ,
\end{align}
where $\hat{a}$ is the lowering ladder operator of the unshifted harmonic oscillator and $\lambda$ specifies how far the state is shifted in oscillator units, i.e. $\langle x|\hat{U}_{\mathrm{shifted}}|n\rangle=\phi_n\left(x+\lambda\right)$ for any harmonic oscillator eigenfunction $\langle x|n\rangle=\phi_n\left(x\right)$.  We can re-express the action of the lowering operator in mode space, $\hat{a}|n\rangle=\sqrt{n}|n-1\rangle$, in terms of energy lattice operators as
\begin{align}
\label{eq:laddertoenergyspace}\hat{a}&\to\sum_j \sqrt{j}\hat{b}_{j-1}^{\dagger}\hat{b}_j\, .
\end{align}
With this, the COBH for a shifted harmonic trap is
\begin{align}
\hat{H}_{\mathrm{shifted}}&=i\log \hat{U}_{\mathrm{shifted}}\, ,\\
\label{eq:shiftedCOBH}&=i\sum_{j\sigma}\left[\sqrt{\frac{j}{2}} \lambda_{\sigma }\hat{b}_{j-1,\sigma}^{\dagger}\hat{b}_{j,\sigma}-\mathrm{H.c.}\right]\, ,
\end{align}
written in terms of the annihilation operators of the energy lattice $\hat{b}_{j,\sigma}$ for site $j$ and spin $\sigma$ and the spin-dependent displacements $\lambda_{\sigma}$.  This COBH is a tight binding model with pure imaginary hopping coefficients varying as $\sqrt{j}$ with lattice site.  

In the case of a change in trap frequency, the transformation is 
\begin{align}
\hat{U}_{\mathrm{dilated}}&=e^{\log \lambda \left(\hat{a}^2-(\hat{a}^{\dagger})^2\right)/2}\, ,
\end{align}
where $\lambda$ is the scale parameter such that $\langle x|\hat{U}_{\mathrm{dilated}}|n\rangle= \sqrt{\lambda} \phi_n\left(\lambda x\right)$.  Expressed in terms of the new trapping frequency $\tilde{\omega}$, $\lambda=\sqrt{{\tilde{\omega}}/{\omega}}$.  The COBH for a general spin-dependent change of trapping frequency $\omega\to\omega_{\sigma}$ is hence 
\begin{align}
\hat{H}_{\mathrm{dilated}}&=i\log\hat{U}_{\mathrm{dilated}}\, ,\\
\label{eq:dilatedCOBH}&=i\sum_{j\sigma}\left[\frac{\sqrt{j\left(j-1\right)}\log \omega_{\sigma}/\omega}{4} \hat{b}_{j-2,\sigma}^{\dagger}\hat{b}_{j,\sigma}-\mathrm{H.c.}\right]\, ,
\end{align}
where we have again mapped mode operators to the energy lattice using Eq.~\eqref{eq:laddertoenergyspace}.  Eq.~\eqref{eq:dilatedCOBH} is a model involving only next-nearest neighbor hopping with pure imaginary hopping coefficients.  

In the general case, one can find the COBH numerically as ${H}=i\log {U}$, with ${U}$ defined in Eq.~\eqref{eq:Unntilde}.  Unitary operators are normal, and so an appealing method to find the logarithm is to compute the spectral decomposition using numerical eigensolver routines.  Standard eigensolver approaches for general (i.e.~non-symmetric) matrices, such as ZGEEV in Lapack~\cite{Lapack}, cannot guarantee orthogonality of the eigenvectors when eigenvalues are near-degenerate.  Instead, it is desirable to compute the complex-valued Schur decomposition ${U}={Q}{R} {Q}^{\dagger}$ with ${Q}$ unitary and ${R}$ upper triangular.  Routines for computing this form, such as ZGEES, do guarantee orthogonality of the vectors in ${Q}$.  Further, since ${U}$ is normal, ${R}$ must be diagonal with the eigenvalues as entries, and so ${H}_{n \tilde{n}}=i\sum_m {Q}_{n m}\log \left(R_{mm}\right){Q}_{\tilde{n} m}^{\star}$.  As mentioned above, the transformation matrix Eq.~\eqref{eq:Unntilde} is generally not unitary for a finite-dimensional set of states.  Even if this operator is ``approximately unitary" for some subset of states $\{|q\rangle \}$ in the sense that $\sum_{m} U_{qm}U_{qm}^{\star}\approx 1$ for these states, the Schur form of this operator will be non-diagonal due the other states not in this set.  Hence, it is essential as an intermediate step to compute the unitary matrix nearest to ${U}$ in the least-squares sense, which is obtained as ${U}=\mathcal{U}\mathcal{V}^{\dagger}$, where $\mathcal{U}\Sigma \mathcal{V}^{\dagger}$ is the singular value decomposition of ${U}$, and use this matrix as input for the Schur decomposition.  While this operator will not act appropriately on the entire set of basis states, it will reproduce the appropriate unitary action on the states $\{|n\rangle\}$ which were near-unitary in the sense defined above.  Hence, by increasing the basis size, we can construct the correct unitary transformation on any desired subset of the basis states.

 \begin{figure}[t]
\includegraphics[width=.99\columnwidth,angle=0]{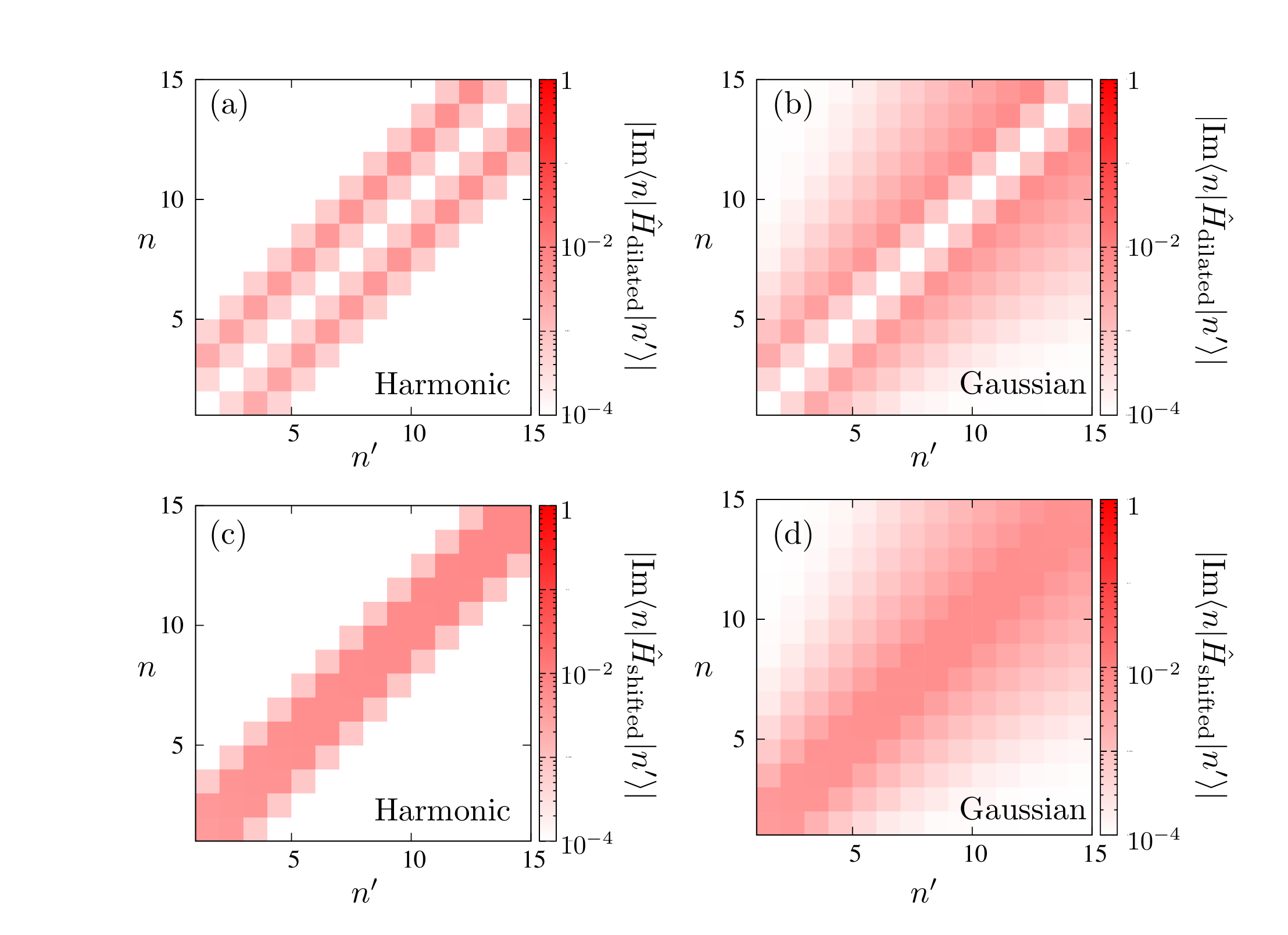}
\caption{(color online) \emph{Change of basis Hamiltonians for shifted and dilated traps.}  The absolute magnitude of the matrix elements of the change of basis Hamiltonian (COBH) $H$ are shown versus eigenstate index.  The top panels (a)-(b) are for dilated traps and the bottom panels (c)-(d) for shifted traps.  The left panels (a), (c) show the results for a harmonic oscillator, demonstrating the strictly banded structure of Eqs.~\eqref{eq:shiftedCOBH} and \eqref{eq:dilatedCOBH}.  The right panels (b), (d) show the results for a deep Gaussian trap, in which anharmonicity smears out the banded structure.  The smearing is greater for higher eigenstates, which sample more of the anharmonic regions of the trap.
 \label{fig:COBH}}
\end{figure}

Examples of the COBH matrix elements $H_{nn'}$ are given in Fig.~\ref{fig:COBH} for the cases of a shifted trap and a dilated trap, using a trap displacement of $0.2a_H$, $a_H$ the harmonic oscillator length, and an increase in the trapping frequency by $0.1\omega$, respectively.  The left panels show the results for a harmonic oscillator, and display the band-diagonal structure specified above with matrix elements increasing with $n$.  The right panels show the numerically obtained COBH for a 1D Gaussian trap of the form $Ve^{-2x^2/\ell^2}$.  Here, we take the depth $V/E_{\ell}=800$, where $E_{\ell}=\hbar^2/(2m\ell^2)$ is the energy associated with the trap length scale $\ell$.  The shift and increase in trap frequency used in the COBH calculation use the harmonic approximations obtained by a quadratic expansion near the trap minimum: $\hbar\omega=\sqrt{8E_{\ell}V}$ and $a_H=\ell/(2V/E_{\ell})^{1/4}$.  The plots show the expected behavior: the dominant COBH structure of the Gaussian trap is the same as the corresponding harmonic trap.  However, there are more nonzero elements, and the magnitudes of these elements increase as the eigenstate energy increases, as these eigenstates sample more of the anharmonic regions of the trap.

\subsection{Representation of Inhomogeneous long-range interactions}
\label{sec:RepLRI}
An important intermediate step in applying MPS methods to a model with long-range interactions is to write the Hamiltonian as a matrix product operator.  For a translationally invariant, decaying interaction of the form
\begin{align}
\sum_{i<j} f\left(j-i\right)\hat{A}_i\hat{B}_j\, ,
\end{align}
a well-established method exists, known as the Hankel singular value decomposition, which amounts to least-squares fitting the interaction function $f\left(r\right)$ by a sum of (possibly complex) exponentials~\cite{Murg,Crosswhite}.  That is, the ansatz
\begin{align}
\label{eq:TIf} \tilde{f}\left(r\right)&=\sum_{n=1}^{n_{\mathrm{exp}}} J_n \lambda_n^{r}\, ,
\end{align}
is optimized via the functional $\sum_{r=1}^{r_{\mathrm{max}}}\left|f\left(r\right)-\tilde{f}\left(r\right)\right|^2$ with the maximum range $r_{\mathrm{max}}$ and the number of exponentials $n_{\mathrm{exp}}$ as convergence parameters.

This approach no longer holds for interactions which are not translationally invariant.  While interactions represented in real space are translationally invariant, interactions on the energy lattice are strongly inhomogeneous (see Fig.~\ref{fig:Interactions}), and so alternative methods are required.  Non-translationally-invariant interactions also appear in other contexts, such as quantum chemical calculations with DMRG~\cite{qchemMPO} and in the representation of 2D systems$-$even systems with translationally invariant interactions$-$by mapping to a 1D chain~\cite{Stoudenmire_2D}.  

The most direct method of constructing an MPO representation is to variationally optimize the elements of the MPO itself, which can be done using the methods developed for MPSs by mapping this operator to a state on a lattice with $d_i^2$ dimensional local Hilbert spaces.  While these methods have been successfully used in some cases~\cite{PhysRevA.81.062337}, they suffer from potential drawbacks.  First, while MPO constructions of known Hamiltonian are extremely sparse, this variational method does not preserve sparsity and in general produces fully dense MPOs (though this can be alleviated in some cases, see, e.g.~\cite{PhysRevB.95.035129}).  Even in the case that the bond dimension of the MPO, which is often used as a proxy for its complexity, is small, the total number of operations required to apply this MPO to an MPS may be larger than for a sparse MPO with larger bond dimension.  In addition, variational fitting of MPOs can be numerically costly, amounting to an MPS simulation with high-dimensional local Hilbert spaces.  This optimization must be performed whenever any change is made to the operator, and so previous optimizations or results cannot be re-used easily.

Here, we propose an alternative approach, in which the parameters appearing in the exponential fit, Eq.~\eqref{eq:TIf}, are allowed to vary in space.  Namely, given a generic interaction of the form $\sum_{i<j} f\left(i,j\right)\hat{A}_i\hat{B}_j$, we fit the terms with $i<j$ with the ansatz
\begin{align}
\label{eq:NTIf} \tilde{f}\left(i,j\right)&=\sum_{n=1}^{n_{\mathrm{exp}}} J_{i,n}\prod_{k=i}^{j-1}\lambda_{k,n}\, .
\end{align}
The ansatz Eq.~\eqref{eq:NTIf} is a natural extension of Eq.~\eqref{eq:TIf} in which the coupling constants $\{\mathbf{J}_i\}$ and decay parameters $\{\boldsymbol{\lambda}_k\}$ are allowed to vary along the chain.  To find these site-dependent coefficients, we solve the least-squares minimization problem
\begin{align}
\label{eq:nonTILS}\min_{\mathbf{J}_1\dots \mathbf{J}_L; \boldsymbol{\lambda}_1\dots \boldsymbol{\lambda}_{L-1}}\sum_{p=1}^{L-1}\sum_{i=1}^{L-p}\left|f\left(i,i+p\right)-\sum_{n=1}^{n_{\mathrm{exp}}} J_{i,n}\prod_{j=i}^{i+p-1}\lambda_{j,n}\right|^2\, .
\end{align}
 A direct non-linear least squares minimization of Eq.~\eqref{eq:nonTILS} using the Levenberg-Marquardt algorithm~\cite{marquardt1963algorithm} has proven to be ill-conditioned due to strong cross-correlations between variations in the $\lambda$ parameters.  Using standard regularization procedures, e.g., Tikhonov or SVD regularization~\cite{press1993}, to improve the conditioning has the unfortunate consequence of introducing an imaginary component to $\tilde{f}$ on the order of the regularization parameter.

To avoid the above drawbacks of direct non-linear least-squares minimization of Eq.~\eqref{eq:nonTILS}, we instead use an alternating least-squares approach which is very similar to the sweeping method used in DMRG~\cite{MurgReview,Schollwoeck}.  The motivation for this is the fact that the functional Eq.~\eqref{eq:nonTILS} with all parameters held fixed except for one (either a $\boldsymbol{\lambda}_j$ or a $\mathbf{J}_j$) is a quadratic form in that parameter.  For example, we can write the functional as a quadratic form in $\boldsymbol{\lambda}_q$ as
\begin{widetext}
\begin{align}
\nonumber &\sum_{p=1}^{L-1}\sum_{i=1}^{L-p}\Big\{\left|f\left(i,i+p\right)\right|^2-[i\le q\le i+p-1]f\left(i,i+p\right){\mathbf{J}}^{\dagger}_{i}{\prod_{j=i}^{i+p-1}}'{\Lambda}_{j}^{\star}{\boldsymbol{\lambda}}_q^{\star}-[i\le q\le i+p-1]f^{\star}\left(i,i+p\right)\mathbf{J}^{T}_{i}{\prod_{j=i}^{i+p-1}}'\Lambda_{j}{\boldsymbol{\lambda}}_q\\
&+\left({\boldsymbol{\lambda}}_q^{\dagger}{\prod_{j=i}^{i+p-1}}'{\Lambda}_{j}^{\star}{\mathbf{J}}_i^{\star}\right)\left(\mathbf{J}^{T}_i {\prod_{j'=i}^{i+p-1}}' \Lambda_{j}\boldsymbol{\lambda}_q\right)\Big\}\, ,
\end{align}
\end{widetext}
where the prime on the products indicates that $j=q$ is omitted, $\Lambda$ denotes the diagonal matrix with $\boldsymbol{\lambda}$ as entries, and the Iverson bracket $\left[a\right]=1$ if $a$ is true and 0 otherwise.  Taking the derivative with respect to $\boldsymbol{{\lambda}}_q^{\star}$ and setting it equal to zero, this quadratic form defines a linear system of equations for minimization
\begin{widetext}
\begin{align}
 \left(\sum_{p=1}^{L-1}\sum_{i=1}^{L-p}[i\le q\le i+p-1] {\prod_{j=i}^{i+p-1}}'{\Lambda}_{j}^{\star}{\mathbf{J}}_i^{\star}\mathbf{J}^{\dagger}_i {\prod_{j'=i}^{i+p-1}}' \Lambda_{j}\right)\boldsymbol{\lambda}_q&=\sum_{p=1}^{L-1}\sum_{i=1}^{L-p}[i\le q\le i+p-1]f\left(i,i+p\right){\mathbf{J}}^{\dagger}_{i}{\prod_{j=i}^{i+p-1}}'{\Lambda}_{j}^{\star}\, .
\end{align}
\end{widetext}
The right-hand side is a constant vector, and parentheses on the left-hand side denotes the matrix of correlations between changes in $\boldsymbol{\lambda}_q$ and all other parameters.  Solving this equation in the least-squares sense, i.e.~applying the pseudoinverse of the matrix on the left hand side to the right hand side, gives the optimal exponential decay parameters $\boldsymbol{\lambda}_q$ for minimizing our functional with all other parameters held fixed.  We will refer to the minimization of our functional with respect to a single parameter with all others held fixed as \emph{local} minimization.  This is exactly in analogy with the local minimization in variational MPS algorithms~\cite{Schollwoeck}, in which the matrices corresponding to a single site are optimized with all others held fixed.  The local minimization with respect to $\mathbf{J}_i$ is given by the linear system of equations
\begin{align}
\nonumber & \left(\sum_{p=1}^{L-i}{\prod_{j=i}^{i+p-1}}{\Lambda}_{j}^{\star}{\prod_{j'=i}^{i+p-1}} \Lambda_{j}\right)\mathbf{J}_q\\
&=\sum_{p=1}^{L-i}[i\le q\le i+p-1]f\left(i,i+p\right){\boldsymbol{\lambda}}_i^{\dagger}{\prod_{j=i+1}^{i+p-1}}{\Lambda}_{j}^{\star}\, .
\end{align}
The complete procedure is to ``sweep" across parameters, performing local minimization on the $\lambda$s and $J$s in turn, until convergence is reached.  This algorithm is much more stable than a direct global least-squares minimization, and also produces real approximations $\tilde{f}$.

As our proposed scheme is essentially variational, the quality of our final state and the rate of convergence are greatly improved by having a good initial guess.  We do so by finding the nearest translationally invariant interaction, and then fitting that interaction using the Hankel singular value decomposition.  Finding the nearest translationally invariant interaction is equivalent to finding the nearest matrix to $f$, denoted $f_T$, which has constant diagonals (i.e.~a Toeplitz matrix).  This is
\begin{align}
\label{eq:NTI}f_T\left(\left|p\right|\right)&=\frac{\sum_{i=1}^{L-p} f\left(i,i+p\right)}{L-p}\, .
\end{align}
Fitting this with the Hankel singular value decomposition to obtain $\tilde{\mathbf{J}}$ and $\tilde{\boldsymbol{\lambda}}$, we can initialize $\mathbf{J}_i=\tilde{\mathbf{J}}$ and $\boldsymbol{\lambda}_i=\tilde{\boldsymbol{\lambda}}$.  We stress that our method optimizes the coefficient matrix $f$ rather than a many-body operator, and so produces efficient, re-usable approximations for any operators $\hat{A}$ and $\hat{B}$, and does not have to perform any operations in a many-body Hilbert space.

\begin{figure}[t]
\includegraphics[width=.8\columnwidth,angle=0]{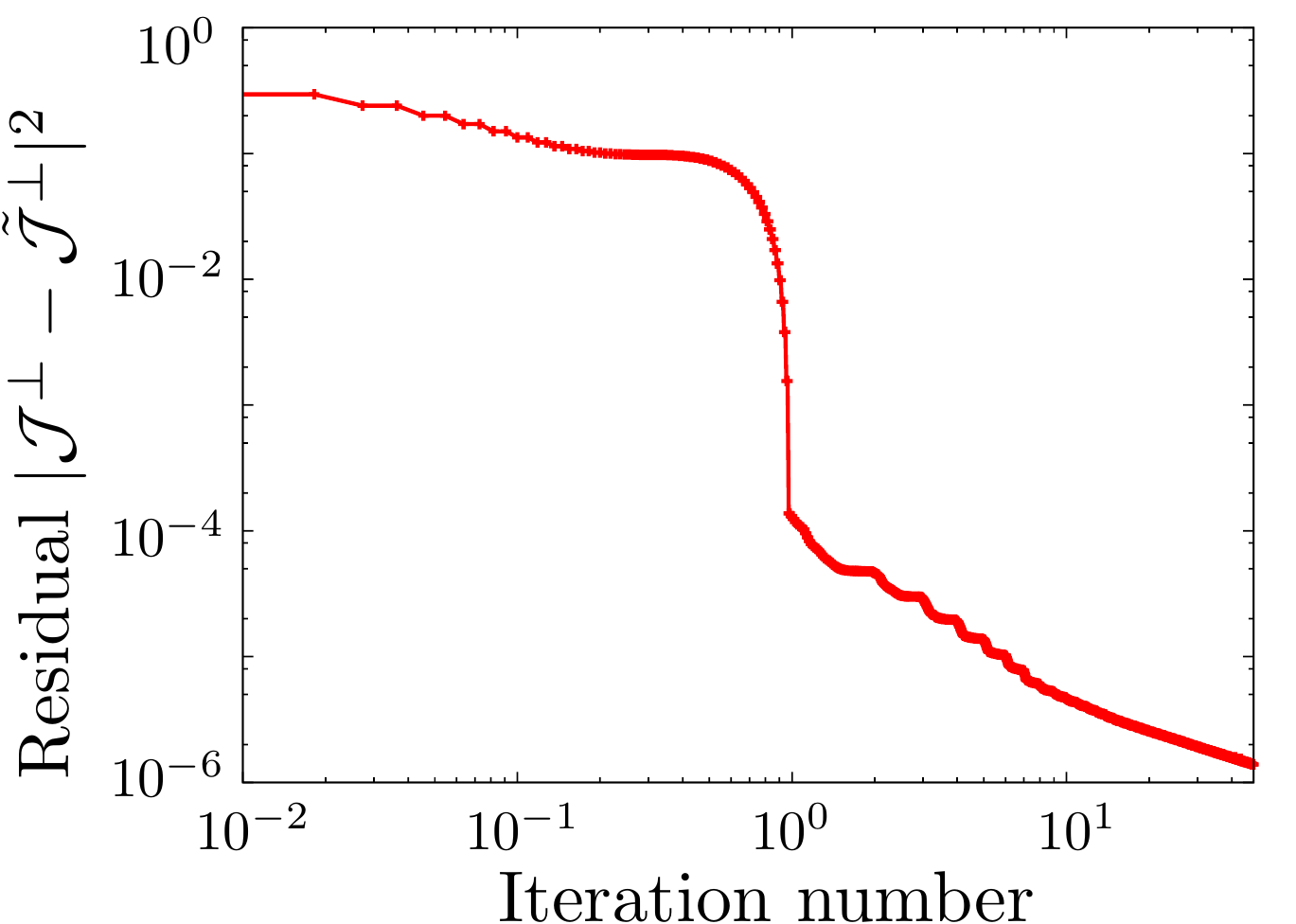}
\caption{(color online) \emph{Exponential Fitting of non-translationally invariant interactions.}  The error of the non-translationally invariant MPO fitting procedure, as quantified by the (Frobenius norm) residual $|\mathcal{J}^{\perp}-\tilde{\mathcal{J}}^{\perp}|^2$ between the exact interaction matrix $\mathcal{J}^{\perp}$ and the approximate matrix $\tilde{\mathcal{J}}^{\perp}$ constructed as in Eq.~\eqref{eq:NTIf}, is shown as a function of the optimization effort, quantified by number of optimization iterations.  Here, one iteration refers to optimization of each $\mathbf{J}_i$ and $\boldsymbol{\lambda}_i$ once.
 \label{fig:BFig}}
\end{figure}

To show how this algorithm works in practice, we will fit the exchange interaction matrix in the case that the particles are subject to a harmonic trap and a magnetic field gradient of strength $m\omega^2x_0^2$~\cite{PhysRevLett.117.195302}, 
\begin{align}
\nonumber \mathcal{J}^{\perp}_{nm}&=\int dx \psi_n\left(x-x_0\right)\psi_n\left(x+x_0\right)\\
&\times \psi_m\left(x-x_0\right)\psi_m\left(x+x_0\right)\, .
\end{align}
 An error-effort plot of the optimization procedure is shown in Fig.~\ref{fig:BFig}.  The nearest-translationally invariant interaction matrix, which is the input to the optimization procedure, has an error $\sim\mathcal{O}\left(1\right)$, while running the optimization procedure for 50 ``sweeps" produces an error nearly six orders of magnitude smaller.  Although our number of variational parameters is much greater compared to the translationally invariant case, the structure and sparsity of the MPO representation have not changed.

\section{Validation of the spin model and importance of corrections}
\label{sec:Validation}

The spin model is, at face value, a rather severe approximation.  In particular, taking the expectation of the spin model Hamiltonian for a non-interacting state is equivalent to first-order perturbation theory in the interaction Hamiltonian.  However, since our approximation is made at the operator level rather than the expectation value level, dynamics obtained with the spin model Hamiltonian includes contributions to all orders in perturbation theory for certain terms in the interaction Hamiltonian, while completely neglecting the effects of other terms.  That is to say, the spin model is not equivalent to perturbation theory in the interaction Hamiltonian for dynamics.  In spite of its seeming severity, the spin model performs extraordinarily well in predicting some quantities, such as the decay of the contrast in Ramsey spectroscopy following a sudden spin-dependent quench of the trapping parameters~\cite{PhysRevLett.117.195302}.  The purpose of this section is to numerically benchmark the spin model against solutions of the full Hamiltonian Eq.~\eqref{eq:FullH} for small systems where computing the full dynamics is possible.  In addition, we consider what the dominant corrections to the spin model are for weak to moderate interactions compared to the trapping frequency, and how the importance of these corrections is modified by adding realistic anharmonicity to the trap.  

 \begin{figure}[t]
\includegraphics[width=.9\columnwidth,angle=0]{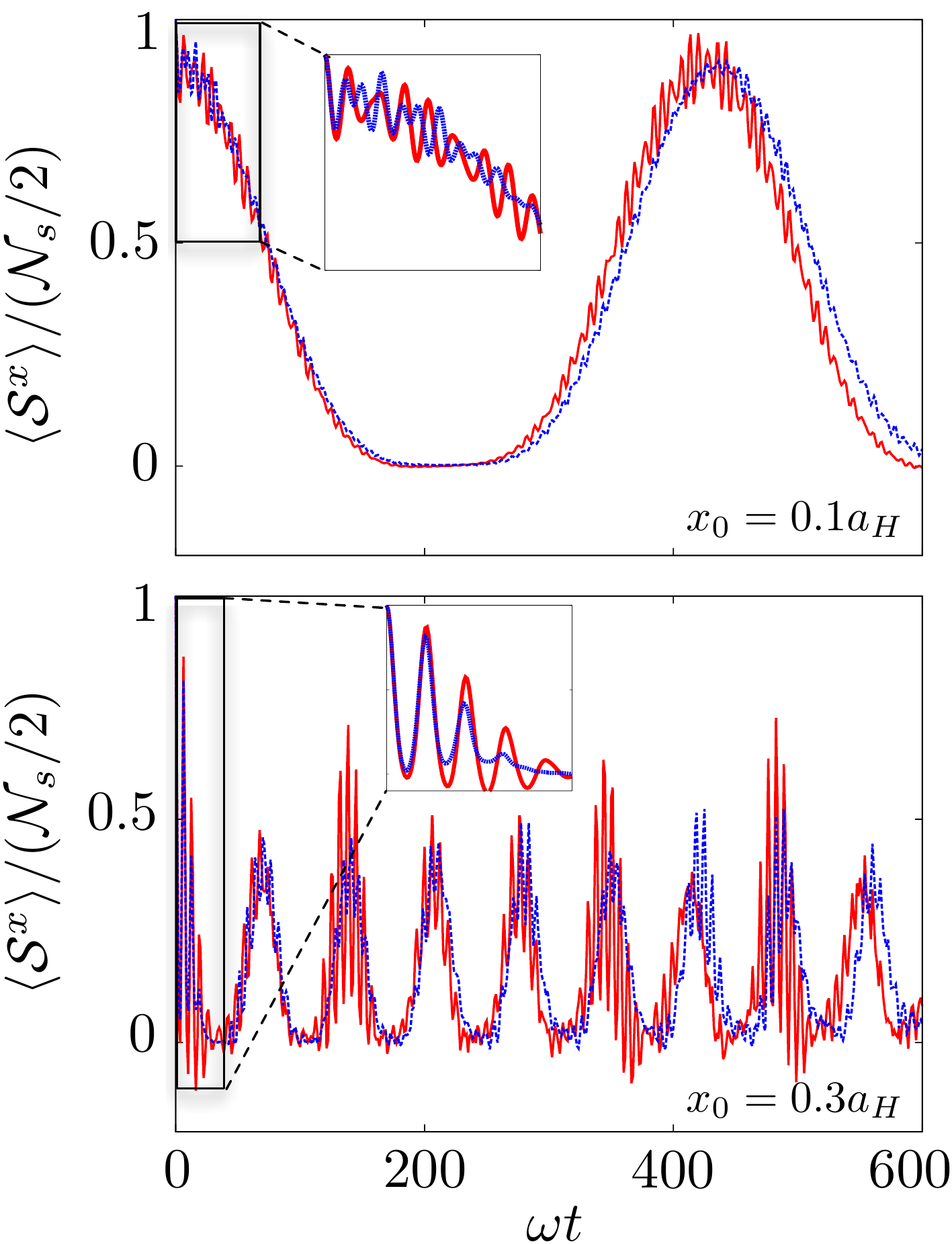}
\caption{(Color online) \emph{Demagnetization following a spin-dependent displacement.}  The exact (red solid) and spin-model (blue dashed) dynamics of the collective spin for an initially polarized state subject to a spin-dependent displacement of harmonic traps by $x_0=0.1a_H$ (top panel) or $x_0=0.3a_H$ (bottom panel) and interaction strength $U=\hbar\omega/\sqrt{8}$.  The overall demagnetization and revival timescales are well-captured by the spin model approach, while smaller-scale features due to interaction-modified motion in the trap are not perfectly captured.  Note that non-interacting motion in the trap is exactly captured within the spin model framework.
 \label{fig:demag}}
\end{figure}
In what follows, we will consider a system with $\ns=5$ spin-1/2 fermions experiencing only $s$-wave interactions.  Our initial state is the motional ground state in a spin-independent harmonic trap with all spins pointing along the $x$ direction.  As all of the particles are prepared identically, they do not experience the $s$-wave interactions.  We then suddenly quench on a spin-dependent potential in the form of a spin-dependent displacement or dilation of the traps.  This change in the trapping potential results in spin-dependent motion in the trap, which in turn leads to $s$-wave collisions.  

One of the most striking consequences of $s$-wave collisions is coherent demagnetization of the system, evinced by decay of $\langle \mathcal{S}^x\rangle$ to zero.  Examples of this demagnetization behavior for an interaction strength $U=\hbar\omega/\sqrt{8}$ and trap displacements of $x_0=0.1a_H$ and $x_0=0.3a_H$ are shown in Fig.~\ref{fig:demag}, with the solutions of the full model corresponding to solid red lines and the spin model prediction as dashed blue lines.  On a coarse scale, we see that the spin model does an excellent job of capturing the overall envelope and timescale of the collapse of magnetization.  However, a closer analysis (insets) shows that there are small oscillations on top of this envelope which correspond to spin-dependent motion in the trap which arises from both single-particle and interaction effects.  The spin model only captures the single-particle component of this motion, and ignores interaction effects, and so only reproduces the exact result at short times.

\begin{figure}[t]
\includegraphics[width=.9\columnwidth,angle=0]{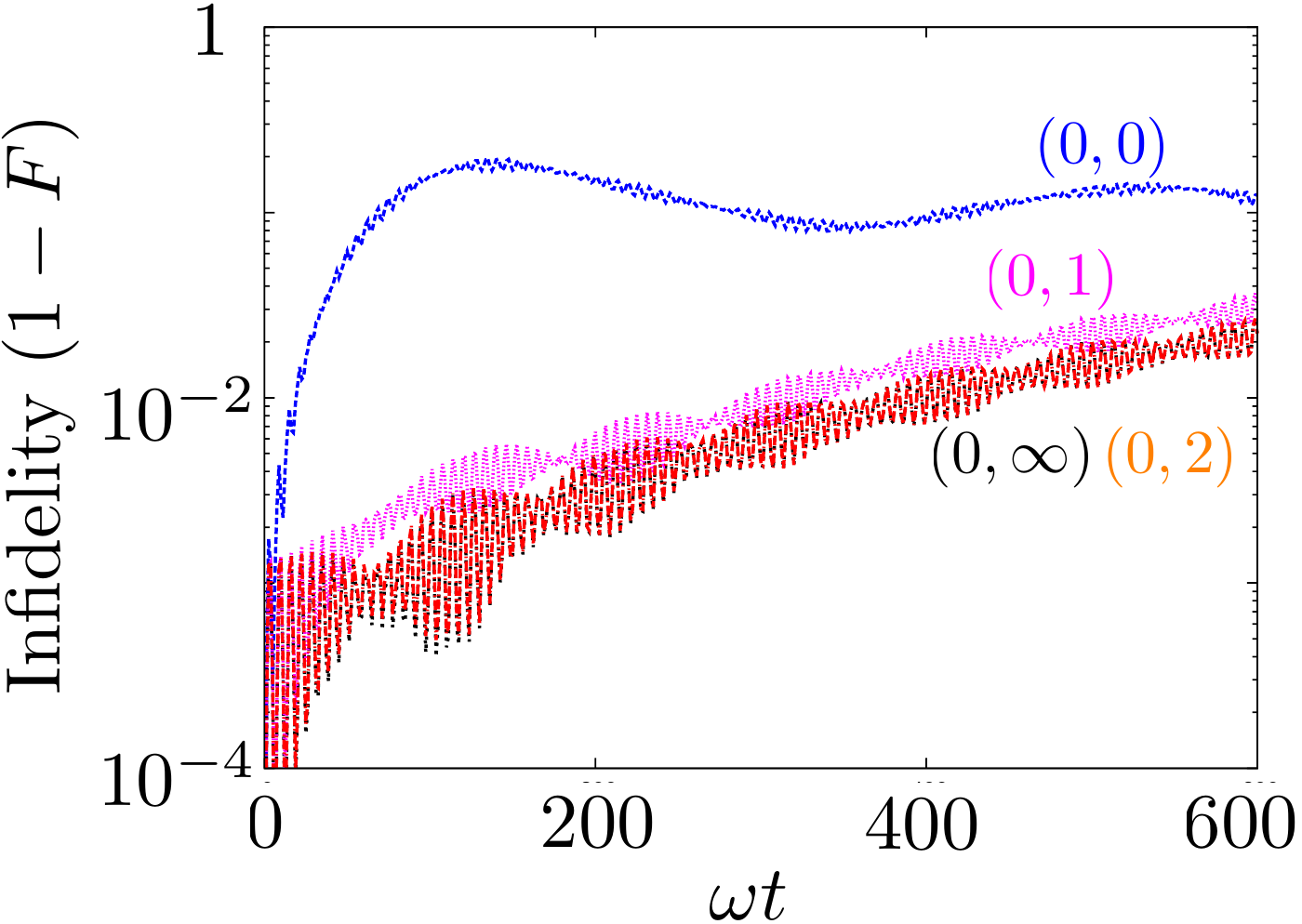}
\caption{(Color online) \emph{Convergence of spin model $(0,d)$ with range $d$.}  Time evolution of the infidelity following a spin-dependent displacement with $x_0=0.1a_H$.  The curves from top to bottom are labeled in terms of increasing mode distance $d$, with the top curve being the spin model result and the bottom curve accounting for all interaction processes that preserve single-particle energy.  Results converge rapidly with increasing $d$ in this case.  
 \label{fig:InfidelityConv}}
\end{figure}

The full model Eq.~\eqref{eq:FullH}, which fully incorporates the effects of interactions on motion in the trap, contains in principle $\mathcal{O}\left(N_{\mathrm{modes}}^4\right)$ parameters for a fixed set of $N_{\mathrm{modes}}$ single-particle modes.  In contrast, the spin model contains only $\mathcal{O}\left(N_{\mathrm{modes}}^2\right)$ parameters.  Clearly, not all of the neglected parameters contribute equally, and so a natural question is which of these parameters are most important for capturing interaction effects on motion.  Motivated by perturbation theory, we expect that the parameters $I_{n_1',n_2'; n_2,n_1}$ which preserve single-particle energy, i.e. $n_1'+n_2'=n_1+n_2$, will be the most relevant, followed by those parameters which change the single-particle energy by $\pm1$ quantum, $\pm2$ quanta, etc.  Further, within the set of parameters which do not change the single-particle energy, those which involve modes $n_1'$ and $n_2'$ which are closest to $n_1$ and $n_2$ (up to exchange) will have a larger matrix element.  Hence, we can classify Hamiltonians including corrections beyond the spin model with two numbers $(\Delta n,d)$, where $\Delta n=\max(n_1'+n_2'-(n_1+n_2))$ is the difference in the single-particle energy of the initial and final configurations and $d$ is the maximum ``mode distance"
\begin{align}
\nonumber \mathcal{M}\left(n_1',n_2';n_1,n_2\right)=\min(&|n_1'-n_1|+|n_2'-n_2|,\\
&|n_1'-n_2|+|n_2'-n_1|)
\end{align}
which accounts for exchange.  In terms of this notation, the spin model is the Hamiltonian $(0,0)$, involving no difference in single-particle energy and also no ``mode separation" between the initial and final configurations.

\begin{figure}[t]
\includegraphics[width=.9\columnwidth,angle=0]{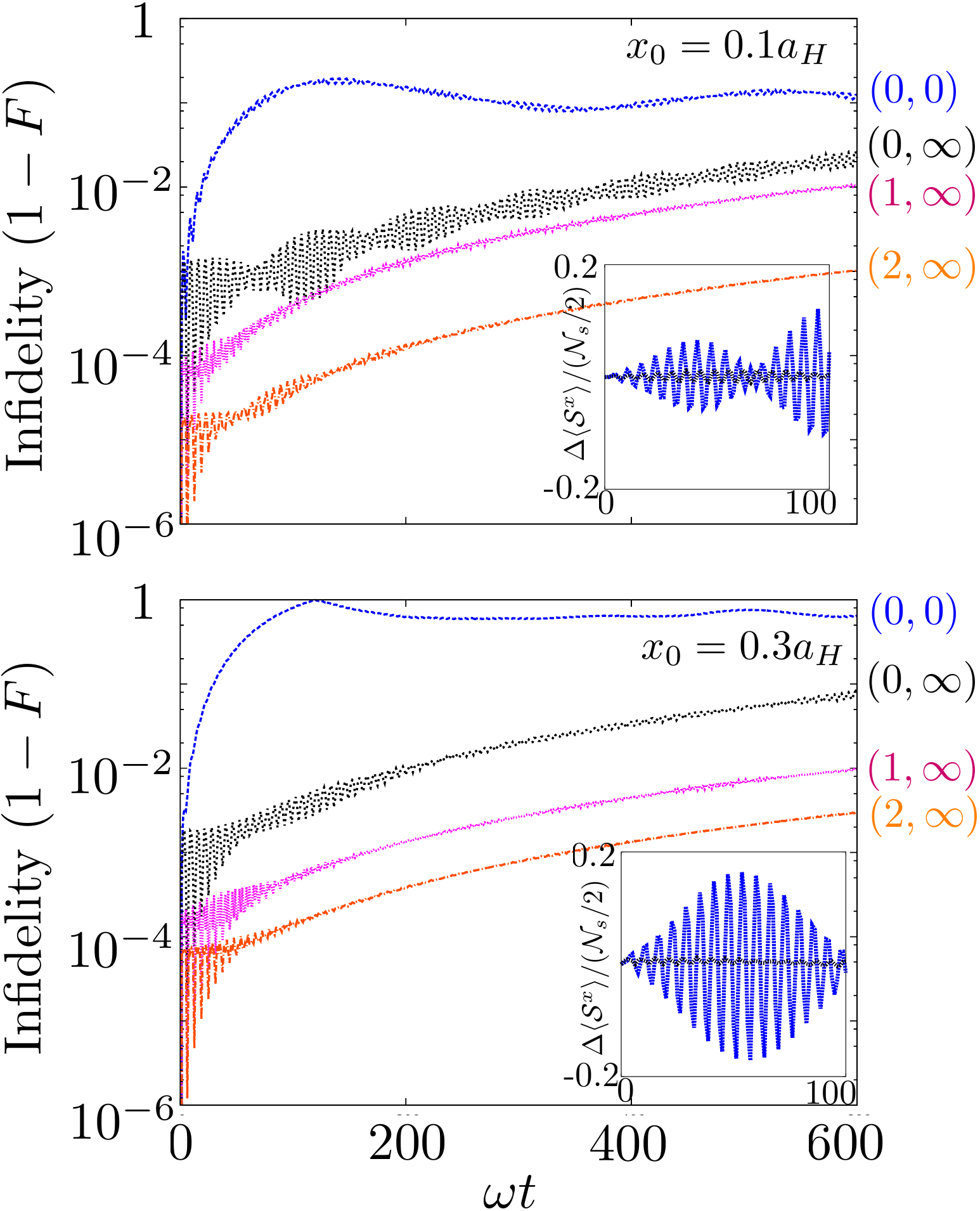}
\caption{(Color online) \emph{Infidelity of spin model in a displaced harmonic trap with and without corrections.}  Time evolution of the infidelity and collective magnetization per spin (inset) following a spin-dependent displacement with $x_0=0.1a_H$ (top panel) and $x_0=0.3a_H$ (bottom panel).  The spin dynamics are well-captured by mode-preserving terms, while the inclusion of mode non-conserving terms increases the fidelity.
 \label{fig:infidelity}}
\end{figure}

As a strict measure of how well the spin model and corrections to it perform, we will consider the fidelity of the state evolved under such an approximate Hamiltonian with the full evolution under the Hamiltonian Eq.~\eqref{eq:FullH}, $F=|\langle \psi_{\mathrm{exact}}|\psi_{\mathrm{approx.}}\rangle|$.  An example for $x_0=0.1a_H$ and $U=0.1\hbar\omega$ is shown in Fig.~\ref{fig:InfidelityConv}.  All curves show models in which single-particle energy is conserved ($\Delta n=0$), while the curves from top to bottom show the results as the mode distance $d$ increases from zero (only mode-preserving direct and exchange terms, i.e.~the spin model) to arbitrary distances.  In this case, the spin model has a roughly 10\% infidelity at long times, while the $d\to\infty$ case improves this to roughly 1\% infidelity.  The infidelity rapidly saturates with $d$ beyond $d=1$ in this case; $d\to\infty$ is essentially indistinguishable from $d=2$ on the scale of this plot.

The top panel of Fig.~\ref{fig:infidelity} again considers the $x_0=0.1a_H$, $U=0.1\hbar\omega$ scenario, but considers two new features.  The first is given in the inset, where we show the error in the collective magnetization per spin for the spin model and $(0,\infty)$ model.  While the spin model has deviations with non-negligible amplitude, we see that the deviations are centered around zero, which is to say that the average, collective behavior is well-captured (see also Fig.~\ref{fig:demag}).  In addition, we see that the $(0,\infty)$ model has spin deviations which are barely perceptible on the scale of this plot.  Hence, single-particle energy-preserving mode changes are the dominant source of the collective spin dynamics.  The second feature of Fig.~\ref{fig:infidelity} is that models with energy non-conserving terms ($\Delta n>0$) are included, and display further improvements in the fidelity.  The bottom panel of Fig.~\ref{fig:infidelity} present the same analysis but for a larger displacement of $x_0=0.3a_H$, and demonstrates that the behavior seen for $x_0=0.1a_H$ is generic.

\begin{figure}[t]
\includegraphics[width=.9\columnwidth,angle=0]{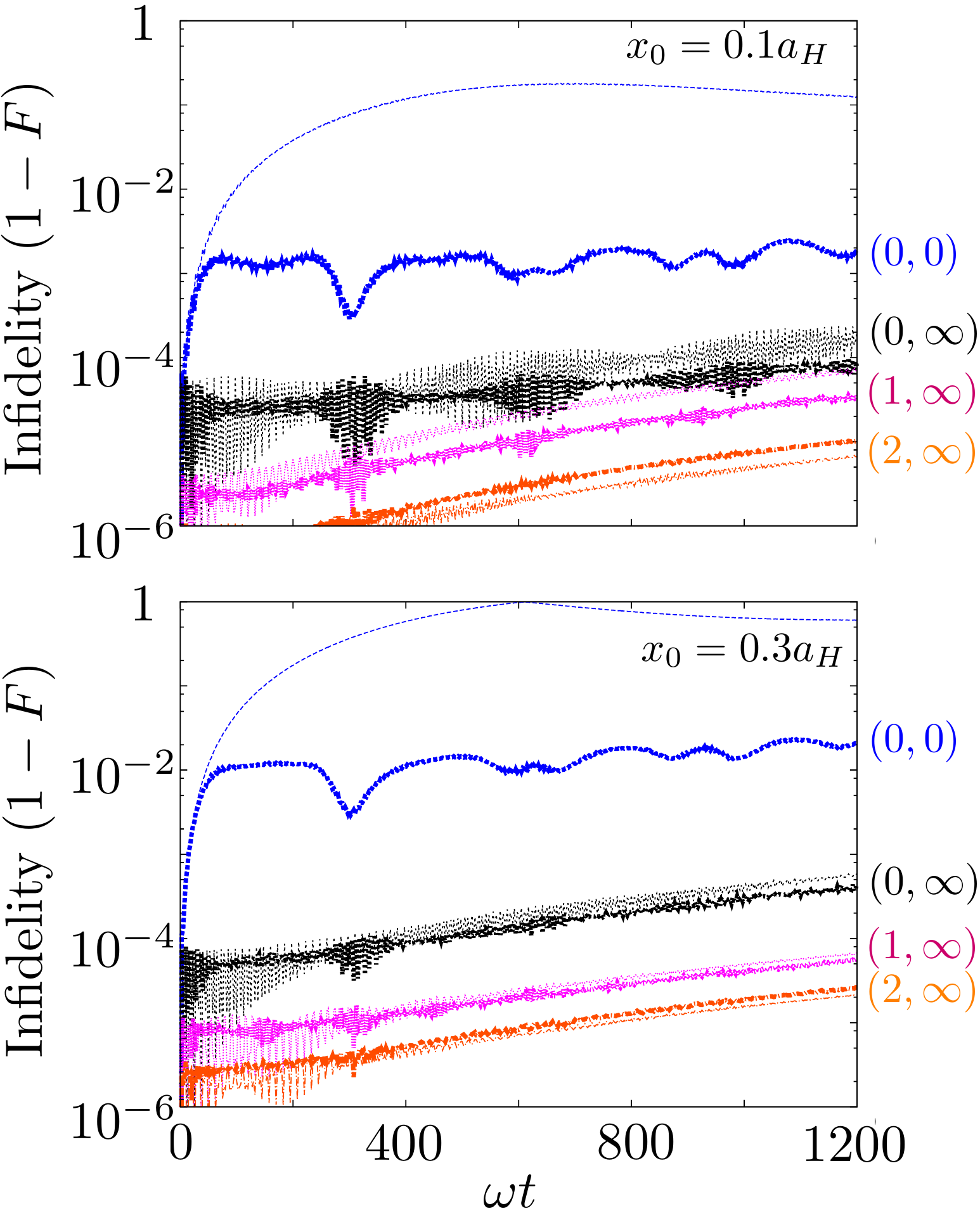}
\caption{(Color online) \emph{Infidelity of spin model in a displaced gaussian trap with and without corrections; weak interactions.} The infidelity of various models in a Gaussian trap (dark lines) compared to a harmonic trap (faint lines), all at $U=0.02\hbar \omega$, which nearly matches the smallest anharmonicity.  Anharmonicity is shown to suppress processes outside the spin model.
 \label{fig:infidelityGaussWeak}}
\end{figure}

From the above, we see that the dominant corrections relevant for capturing the behavior of low-order correlation functions come from single-particle-energy-preserving terms.  In a harmonic potential, there are a large number of such resonances due to the linear spacing of energy levels.  In an anharmonic potential these resonances are no longer exact, and the only exactly resonant collisions are the direct and exchange ones kept by the spin model.  To better understand the effects of anharmonicity on the fidelity of the spin model approach, we consider a Gaussian potential $-V\exp\left(-2x^2/\ell^2\right)$.  The best harmonic approximation to this potential, given by matching the local curvature near $x=0$, yields the harmonic frequency $\hbar\omega =\sqrt{8 E_{\ell} V}$ with $E_l=\hbar^2/(2m\ell^2)$ and the harmonic length $a_H=\ell/(2V/E_{\ell})^{1/4}$.  Treating the quartic-order expansion of the potential in first-order perturbation theory about the harmonic solution, we find the energies
\begin{align}
\label{eq:Eexpan}E_n&\approx -V+\hbar\omega\left(n+\frac{1}{2}\right)+\frac{3}{2}\frac{\hbar\omega}{\sqrt{8\bar{V}}}\left(n^2+n+\frac{1}{2}\right)\, ,
\end{align}
where $\bar{V}=V/E_{\ell}$.  Hence, if we consider an interaction of modes $n$ and $m$ scattering into $(n+d)$ and $(m-d)$, that would be energy conserving in a harmonic trap, this process is off-resonant by an amount
\begin{align}
\Delta E_{n,m,d}&\approx \frac{3 \hbar\omega}{\sqrt{8 \bar{V}}} d\left[d-\left(m-n\right)\right]\, ,\\
&=3E_{\ell}d\left[d-\left(m-n\right)\right]\, .
\end{align}
As expected, this energy difference vanishes for $d=0$ or $d=(m-n)$, corresponding to no change in the modes or a mode swap.

\begin{figure}[t]
\includegraphics[width=.9\columnwidth]{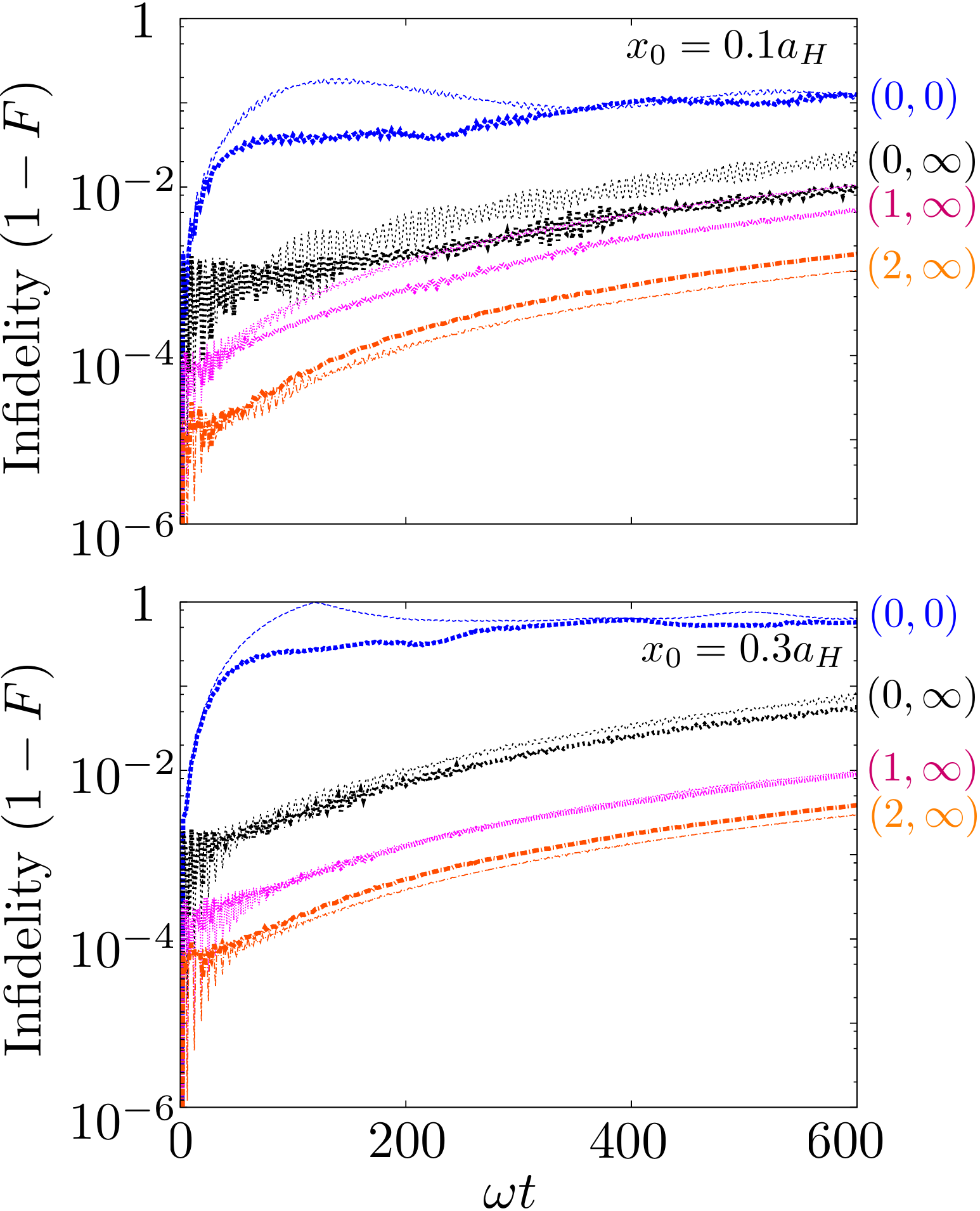}
\caption{(Color online) \emph{Infidelity of spin model in a displaced gaussian trap with and without corrections; strong interactions.} The same as Fig.~\ref{fig:infidelityGaussWeak} but for interaction strength $U=0.1\hbar \omega$ which is nearly 5 times the smallest anharmonicity.  Here, the results for the Gaussian and harmonic traps behave similarly.
 \label{fig:infidelityGauss}}
\end{figure}

For the purposes of understanding the relative impact of anharmonicity on the fidelity of the spin model, it would be most useful to work at fixed interactions and particle number and modify the anharmonicity through the well depth.  However, there is a direct connection between the depth of the well, and hence the anharmonicity, and the number of bound states.  A WKB approximation for the number of bound states yields
\begin{align}
N_b&\approx \sqrt{\frac{\bar{V}}{\pi}}+\frac{1}{2}\, .
\end{align}
Hence, for fixed $\bar{U}=U/\hbar\omega$, the ratio of anharmonicity to interactions scales as
\begin{align}
\frac{\Delta E_{n,m,d}}{\bar{U}\hbar\omega }&\approx \frac{3}{\bar{U}\sqrt{8 \pi}\left(N_b-1/2\right)} d\left[d-\left(m-n\right)\right]\, ,
\end{align}
which shows that the ratio can only be changed by changing $U$ at fixed number of bound states.  Taking $N_b\sim30$ bound states gives $\bar{V}\sim 3000$, and so for the interactions to be on the same order as the anharmonicity requires ${U}\approx 0.02\hbar \omega$.  Fig.~\ref{fig:infidelityGaussWeak} shows the analog of Fig.~\ref{fig:infidelity} in a Gaussian trap with $\bar{V}=3000$ and $U=0.02\hbar\omega$, where at $t=0$ the two spin states feel potentials $-V\exp\left(-2\left(x\pm x_0\right)^2/\ell^2\right)$.  That is to say, the potential itself is shifted rather than a constant gradient applied (recall the two operations are identical for a harmonic trap).  The dark lines are the results for the Gaussian trap, and the faint lines are the corresponding harmonic trap results.  We see a very marked increase in the fidelity of the spin model for the Gaussian trap compared to the harmonic well by more than an order of magnitude on average over the times considered, with more modest gains for the models that include $\Delta n\ne 0$ or $d\ne 0$.  As interactions are increased relative to the anharmonicity, the Gaussian and harmonic oscillator results again become comparable, as shown in Fig.~\ref{fig:infidelityGauss} for $U=0.1\hbar\omega$, roughly 5 times the anharmonicity.  Similar conclusions are expected to hold in higher dimensions, where the density of degenerate harmonic oscillator modes grows more rapidly.  We note that current optical lattice clocks operate in the regime of interactions $\lesssim$ anharmonicity~\cite{rey14}.

 \begin{figure}[t!]
\includegraphics[width=.9\columnwidth,angle=0]{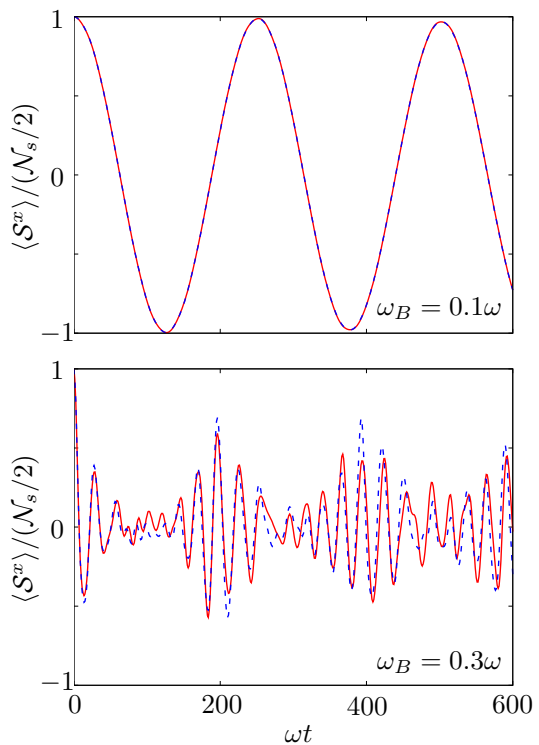}
\caption{(Color online) \emph{Demagnetization following a spin-dependent change in trap frequency.}  The exact (red solid) and spin-model (blue dashed) dynamics of the collective spin for an initially polarized state subject to a spin-dependent change of harmonic trap frequency characterized by $\omega_B=0.1\omega$ (top panel) or $\omega_B=0.3\omega$ (bottom panel) and interaction strength $U=0.4\hbar\omega$.  The overall demagnetization and revival timescales are well-captured by the spin model approach, while smaller-scale features due to interaction-modified motion in the trap are not captured.  Note that non-interacting motion in the trap is exactly captured within the spin model framework.
 \label{fig:Quaddemag}}
\end{figure}

\begin{figure}[t]
\includegraphics[width=.9\columnwidth,angle=0]{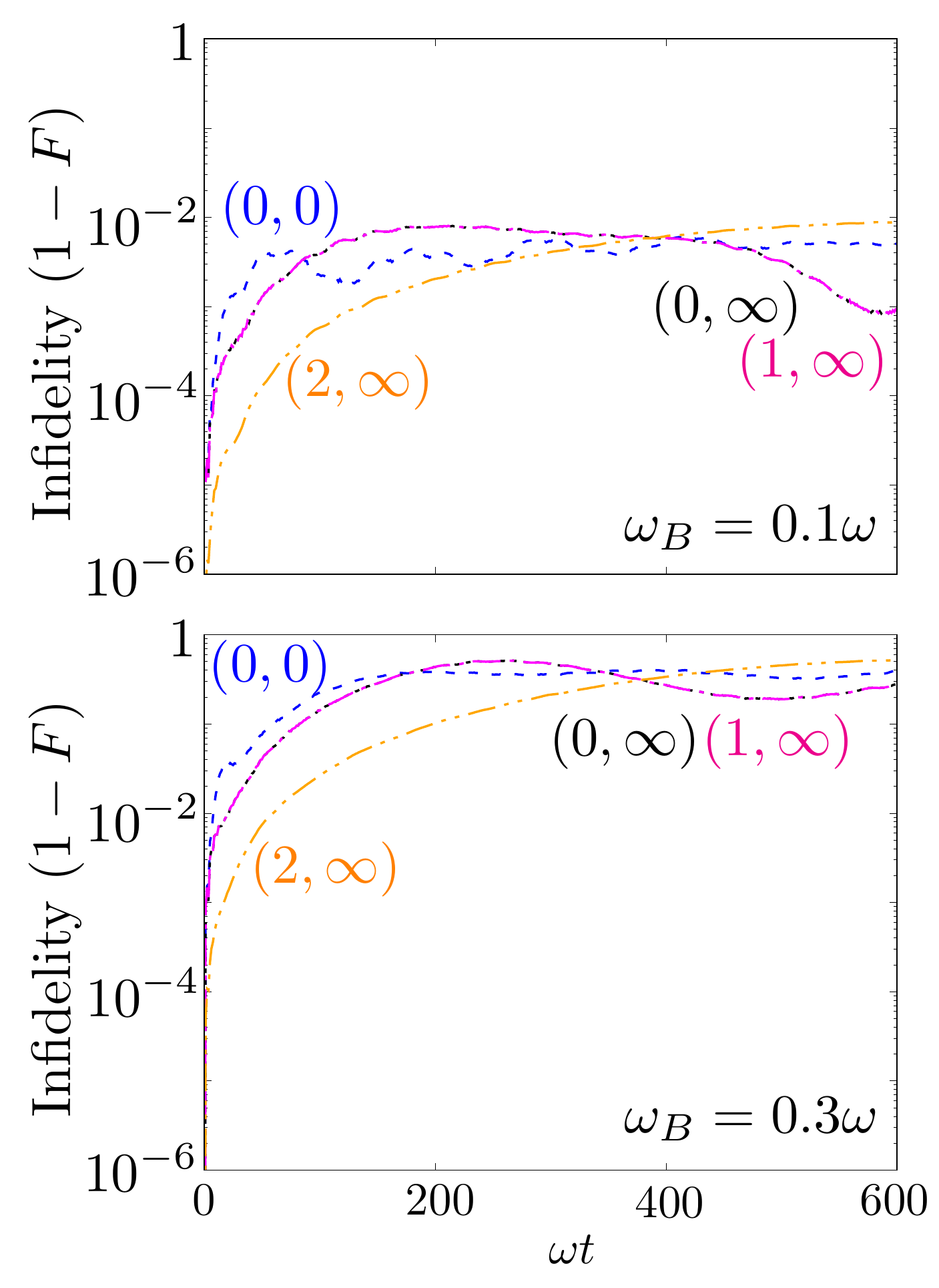}
\caption{(Color online) \emph{Infidelity of spin model in a depth-modulated harmonic trap with and without corrections.}  Time evolution of the infidelity following a spin-dependent dilationwith $\omega_B=0.1\omega$ (top panel) and $\omega_B=0.3\omega$ (bottom panel).  The interaction strength is $U=0.4\hbar\omega$.  The dynamics are captured roughly equally well by the spin model and the considerably more complex $(2,\infty)$ model.
 \label{fig:QuadInfidel}}
\end{figure}

We now turn to the case in which the particles experience a sudden spin-dependent change in trap frequency to new frequencies $\omega_{\sigma}=\sqrt{\omega^2+\sigma \omega_B^2}$.  Here, Fig.~\ref{fig:Quaddemag} is an analog of Fig.~\ref{fig:demag} for the displaced trap case, showing coherent demagnetization due to interactions with weak and strong spin-dependent trapping changes.  As discussed further in Ref.~\cite{PhysRevLett.117.195302}, the top panel is indicative of a regime whose spin dynamics is well-described by a global precession of the collective spin in the XY plane, with the potential inhomogeneity driving oscillations between the manifold of fully collective ``Dicke" states and the neighboring ``spin-wave" manifold.  In this regime, the spin model does an exceptional job of reproducing the exact results.  The bottom panel shows the breakdown of this picture when potential inhomogeneity is increased to become on the same order of the interactions, and deviations of the spin model from the exact result due to interaction-induced mode changes are evident.  This is also formalized through the infidelity in Fig.~\ref{fig:QuadInfidel}, which is analogous to Fig.~\ref{fig:infidelity} for the displaced trap case with larger interactions $U=0.4\hbar\omega$.  In contrast to the displaced case, increasing the mode distance $d$ does not produce a significant decrease in the infidelity; rather, the spin model has qualitatively similar fidelity to the more complex $(2,\infty)$ model.

\begin{figure}[t]
\includegraphics[width=.9\columnwidth,angle=0]{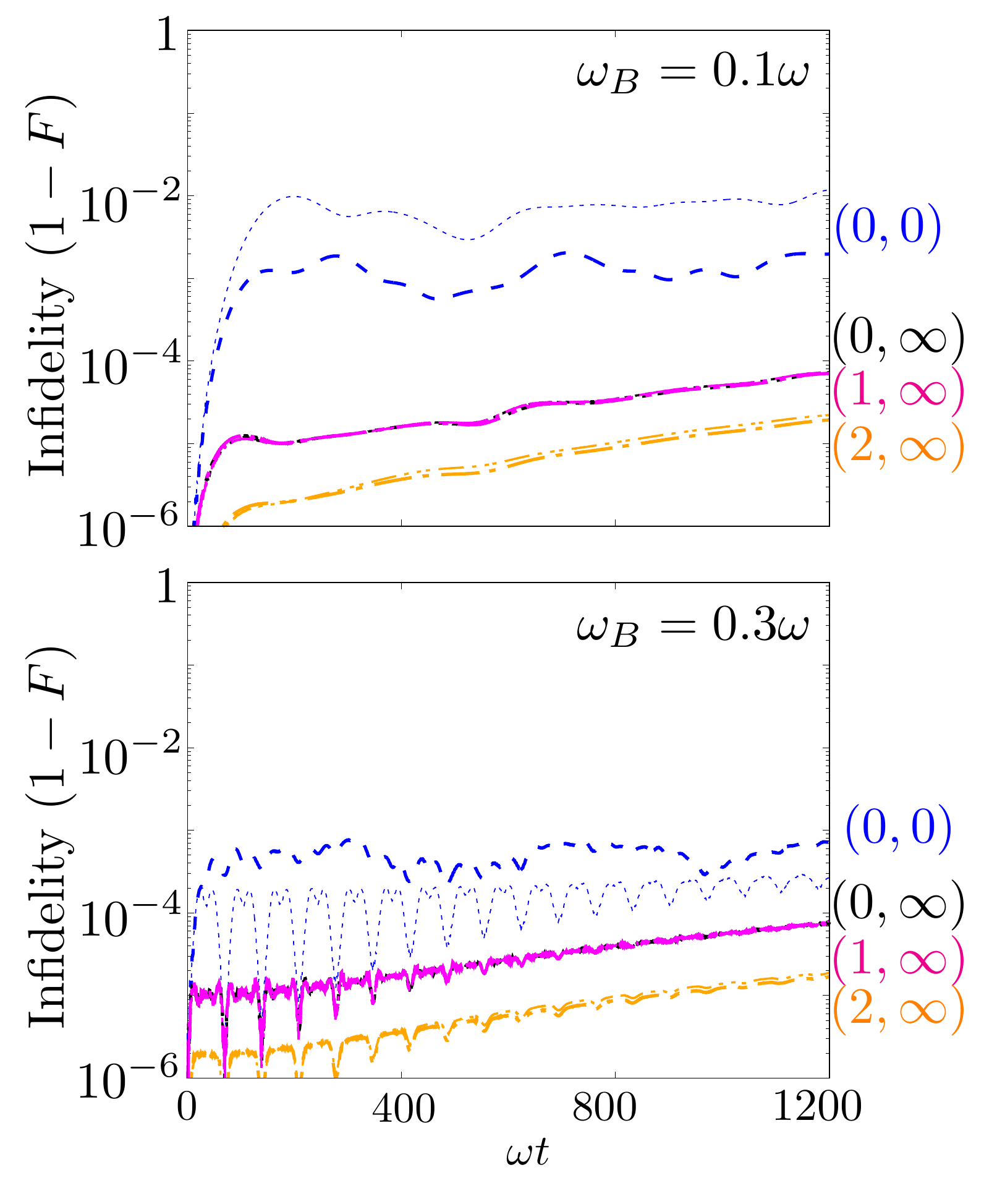}
\caption{(Color online) \emph{Infidelity of spin model in a depth-modulated gaussian trap with and without corrections; weak interactions.} {The infidelity of various models in a Gaussian trap (dark lines) compared to a harmonic trap (faint lines), all at $U=0.02\hbar \omega$, which nearly matches the smallest anharmonicity.  Anharmonicity suppresses processes outside the spin model for small $\omega_B/\omega$ (top panel), but not significantly for larger $\omega_B/\omega$ (bottom panel).}
 \label{fig:QuadGaussInfidelWeak}}
\end{figure}

\begin{figure}[t]
\includegraphics[width=.9\columnwidth,angle=0]{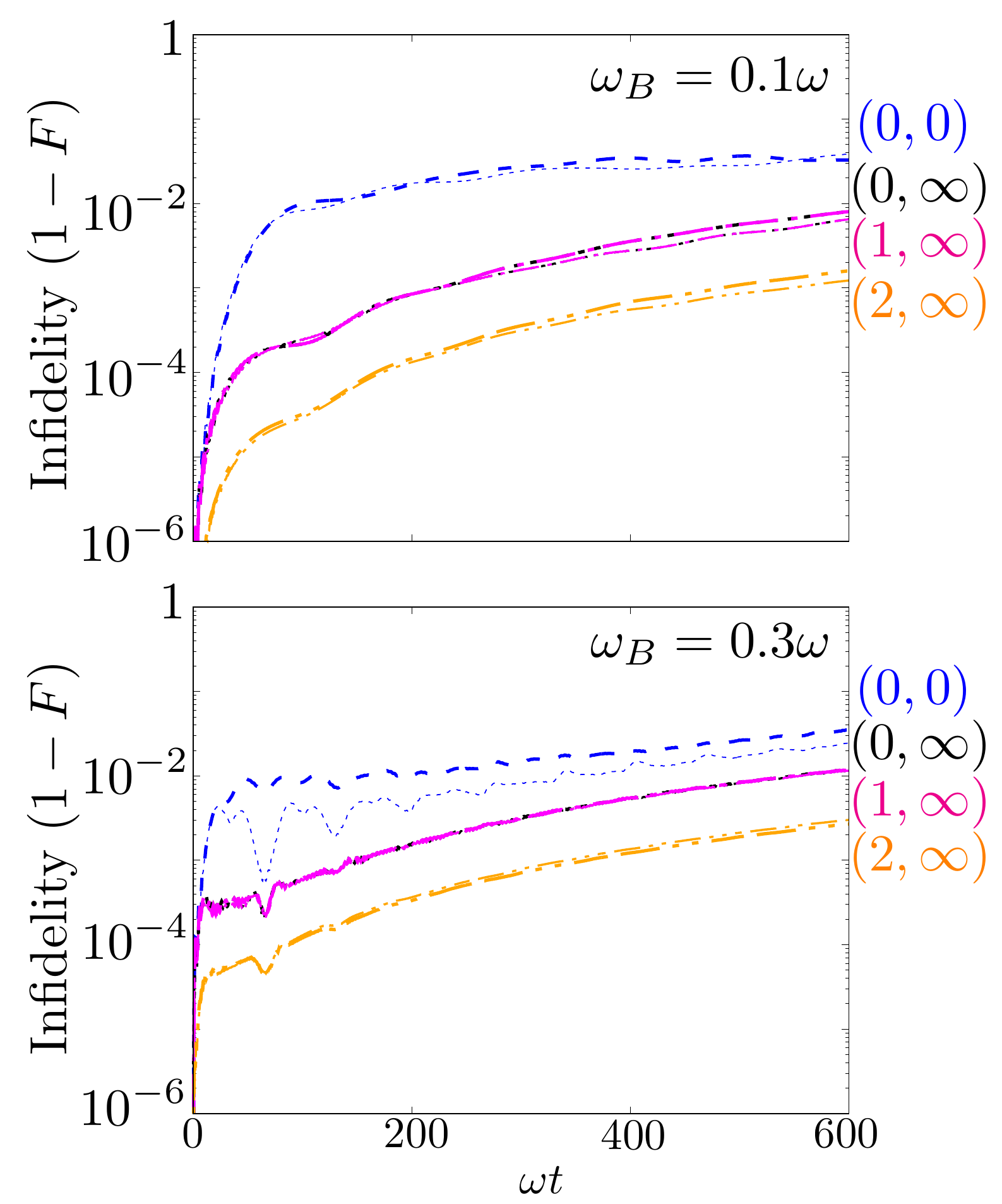}
\caption{(Color online) \emph{Infidelity of spin model in a depth-modulated gaussian trap with and without corrections; strong interactions.} {The infidelity of various models in a Gaussian trap (dark lines) compared to a harmonic trap (faint lines), all at $U=0.1\hbar \omega$, which is roughly five times the smallest anharmonicity.  The harmonic and anharmonic traps behave similarly in this case.}
 \label{fig:QuadGaussInfidelStrong}}
\end{figure}

Finally, we consider the impact of potential anharmonicity on the case of a dilated trap, again considering a Gaussian trap of the form $Ve^{-2x^2/\ell^2}$.  The most natural means of changing the effective trapping frequency would be to change the depth of the potential in a spin-dependent fashion, i.e., $V\to V+\sigma \Delta V$ with $\Delta V=\hbar^2\omega_B^2/(8E_{\ell})=V\omega_B^2/\omega^2$.  However, such a change in depth introduces a homogeneous spin-dependent energy offset of $\sim \Delta V$ that suppresses interaction effects, see Eq.~\eqref{eq:Eexpan}.  Hence, in our simulations, we add a homogeneous potential of $V+\sigma\Delta V$ to best match the spectrum to the harmonic spectrum.  As before, we study the cases of $U$ comparable to the anharmonicity and $U$ large compared to anharmonicity, and take $\bar{V}\sim3000$.  In the former case, shown in Fig.~\ref{fig:QuadGaussInfidelWeak}, where $U=0.02\hbar\omega$, we see that anharmonicity suppresses terms outside of the spin model at small $\omega_B=0.1\omega$.  However, for the case of larger $\omega_B=0.3\omega$ the anharmonic trap result has worse fidelity than the harmonic trap one, though the fidelities of both are quite good, at the $1-10^{-4}$ level.  The case of strong interactions compared to anharmonicity, $U=0.1\hbar\omega$, is shown in Fig.~\ref{fig:QuadGaussInfidelStrong}.  Here, similar to the case of the displaced traps in Fig.~\ref{fig:infidelityGauss}, the harmonic and anharmonic trap results behave similarly.

\section{Conclusions}
\label{sec:Concl}

We discussed the spin model approximation for fermions in spin-dependent potentials, in which interaction processes that change single-particle states are neglected, for both harmonic and anharmonic potentials subject to sudden displacements or changes in depth.  The parameters appearing in these spin models were analyzed for a range of trap displacements and dilations.  Numerical procedures for transforming many-body states between single-particle representations using long-range time evolution under an effective single-particle Hamiltonian and fitting of non-translationally invariant interactions to sums of decaying exponentials were presented, both of which may find other applications in tensor network algorithm applications.  Exact diagonalization simulations were presented for few numbers of particles in various regimes of interaction strength, spin-dependent potential quench strength, and potential anharmonicity, to validate the spin model approximation and understand the importance of the lowest-order corrections beyond it.

I would like to thank Andrew Koller and Ana Maria Rey for collaboration on related topics.

\bibliography{spinsegcitations}

\end{document}